\documentclass[aps,pra,reprint,floatfix,groupedaddress]{revtex4-2}

\usepackage{amsmath, amssymb}
\usepackage{graphicx}
\usepackage{bm}
\usepackage{xcolor}
\usepackage{braket}
\usepackage{placeins}
\usepackage[colorlinks=true,linkcolor=blue,citecolor=blue,urlcolor=blue]{hyperref}
\usepackage{dsfont}
\newcommand{\Ctgt}{C_{\mathrm{target}}}
\newcommand{\Cgen}{C_{\mathrm{gen}}}
\newcommand{\Sx}{\Sigma^{(x)}}

\begin{document}
\raggedbottom

\title{Interpretable Quantum Photonic Generative Models}

\author{Dat Chi Le}
\author{Oleksandr Kyriienko}
\affiliation{School of Mathematical and Physical Sciences, University of Sheffield, Sheffield S10 2TN, United Kingdom}

\date{\today}

\begin{abstract}
We propose a physical generative model for learning correlated multivariate data based on homodyne detection of photonic continuous-variable (CV) circuits. We use an $N$-mode photonic circuit in which two-mode squeezing provides correlations, local phase shifts tune the strength and sign of these correlations, and an adjustable interferometer redistributes them across modes. We train the circuit by matching its Pearson correlation matrix to the one obtained from data. We apply affine post-processing to adjust the marginals independently while keeping the relations intact. We test the model on the multivariate distribution of standardized log returns of financial assets, showing that the photonic model can represent rich correlation structures. We reproduce empirical correlations with residuals below $10^{-4}$ for five variables and recover the correct correlation signs for $94\% \pm 1\%$ of 190 correlation pairs at twenty variables, with $86\%$ of the residuals below $0.1$. As a separate technical contribution, we develop an approach for simulating Kerr gates with small nonlinearity via a Hubbard-Stratonovich expansion, making them suitable for CV-based simulators. Our results establish CV photonic circuits as a platform for interpretable generative learning, suitable for representing and fast sampling of distributions across different application domains.
\end{abstract}

\maketitle

\section{Introduction}
\label{sec:intro}

Many quantitative problems reduce to learning a joint distribution $P(\bm{x})$ of $N$ continuous variables from data and drawing new samples from it \cite{McNeil2005QRM,Goodfellow2016DL}. Generative models solve this problem by capturing the dependence between variables and producing new samples cheaply once trained, as done by variational autoencoders, generative adversarial networks, normalizing flows, and diffusion models \cite{Kingma2014VAE,Goodfellow2014GAN,Dinh2017RealNVP,Papamakarios2021NormalizingFlows,Ho2020DDPM,Song2021ScoreSDE}. However, the sampling cost grows with the data dimension and the model size, as inference requires a pass through the network or a sequence of such passes \cite{Papamakarios2021NormalizingFlows,Song2021ScoreSDE}. This motivates the search for physical systems whose natural measurement statistics realize a trainable multivariate distribution, including coherent Ising machines \cite{McMahon2016CIM,Inagaki2016CIM}, probabilistic bits \cite{Borders2019pbits}, thermodynamic computers \cite{Coles2023ThermodynamicAI}, optical neural networks \cite{Shen2017NanophotonicDL,Wetzstein2020InferencePhotonics}, and polariton condensates \cite{Wang2026PolaritonGAN}.

Quantum generative models implement physical sampling through projective measurements of a parametrized quantum state \cite{Liu2018QCBM,Benedetti2019QML,Tian2023QGMReview}. The Born rule $p(m) = \mathrm{Tr}[\ket{m}\bra{m}\hat{\rho}_{\bm{\theta}}]$ turns each measurement outcome $m$ into a sample from a distribution set by the state $\hat{\rho}_{\bm{\theta}}$ and its circuit parameters $\bm{\theta}$ \cite{Lloyd2018QGAN,Zoufal2019QGAN,Coyle2020BornSupremacy,Du2022PowerQuantumGenerativeLearning}. Models differ in the architecture, the circuit parametrization, the measured observable and its support, and the training strategy \cite{Kyriienko2024TrainableProtocols,Rudolph2024Trainability}. Architectures include qubit-based circuits \cite{Liu2018QCBM,Benedetti2019QML,Coyle2020BornSupremacy,Gili2023Generalization}, photonic samplers based on boson sampling and Gaussian boson sampling \cite{Aaronson2011BosonSampling,Hamilton2017GBS,Kolarovszki2026BosonicBornMachines,Kurkin2026BosonSamplingGenerative}, and fermionic linear-optical samplers \cite{Oszmaniec2022FermionSampling,bako2025fermionic}. Parametrizations include quantum circuit Born machines \cite{Liu2018QCBM,Cheng2018BornVsBoltzmann}, quantum Boltzmann machines \cite{Amin2018QBM,Kieferova2017QBM}, spectral generative models \cite{Kyriienko2024TrainableProtocols,Kasture2023Protocols,Tuysuz2026Fourier,Huang2026SpectralBornMachines}, and models with physically parametrized dynamics \cite{Mahassen2026FokkerPlanck}. Restricted architectures aid classical training while keeping sampling hard, as pursued in ``train-on-classical, deploy-on-quantum'' approaches \cite{RecioArmengol2025TrainClassicalDeployQuantum,Kasture2023Protocols,Lerch2026IQPBornInit,HerreroGonzalez2025BornUltimatum,Huang2025GenerativeQuantumAdvantage}. Training strategies rely on tractable losses, including the maximum mean discrepancy (MMD) \cite{Gretton2012MMD,Liu2018QCBM,Rudolph2024Trainability} and sample-based estimators \cite{Demidik2025SampleBased}.

Photonic systems with continuous-variable (CV) encoding offer a distinct architecture for quantum information processing \cite{Braunstein2005CVReview,Weedbrook2012CVReview,Andersen2015HybridCV}. Optical modes carrying coherent and squeezed light are manipulated with programmable interferometers \cite{Reck1994,Clements2016Clements,Arrazola2021PhotonicML}, and the resulting Gaussian states are described by covariance matrices that are closed-form functions of the circuit parameters, so the correlation structure of the output can be designed. The description leaves the Gaussian family in two cases: when photon-number-resolving detection is used \cite{Hadfield2009SinglePhotonDetectors,Lita2008TES}, and when a non-Gaussian gate based on a Kerr or other nonlinearity is inserted \cite{LloydBraunstein1999,Walschaers2021NonGaussian,Chabaud2023Resources}. The former underlies boson sampling \cite{Aaronson2011BosonSampling,Lund2014GaussianBS,Hamilton2017GBS,Zhong2020Jiuzhang,Madsen2022PhotonicAdvantage}, and the latter underpins non-Gaussian photonic computing \cite{Gottesman2001GKP,Konno2024LogicalStates}. Experimentally, CV systems produce large multimode entangled states \cite{Asavanant2019,Larsen2019}, support multimode homodyne detection at 60 modes \cite{Yoon2026Homodyne}, and can be used to sample from distributions with over more than 400 modes \cite{Cimini2026GBSReservoir}. From the learning perspective, photonic circuits were proposed for quantum neural networks \cite{Killoran2019CVQNN,Steinbrecher2019QONN}, and were recently used to show advantage in detecting signal from limited quantum data \cite{Andersen2025} as well as learning relational information~\cite{Jastrzebski2026RelationalLearning}. Gaussian boson samplers were applied to optimization and graph problems \cite{Arrazola2018MachineLearningPhotonic,Yu2023GBSDrugDiscovery,Wang2025PhotonicsGCN}. For generative modeling, Born machines were built on boson sampling \cite{Salavrakos2024PhotonicBornMachine,Gottlieb2026PhotonicGenerativeTraining,Kolarovszki2026BosonicBornMachines}, and photonic circuits served as generators and latent samplers in adversarial networks \cite{Sedrakyan2024PhotonicQGAN,Bacarreza2025QuantumLatent}. Closest to our work, the CV Born machine of Ref.~\cite{cepaite2022cvbm} reads out homodyne samples of a variational CV circuit and trains it on these samples with a kernel-based loss, with demonstrations on single-mode targets and quantum states of up to three modes. Open questions are whether a structured CV circuit can be trained without sampling, on the covariance of the state alone, such that each optical layer has a stated role in the multivariate correlations of an empirical dataset, and whether this remains trainable at tens of modes.

In this work, we propose a CV photonic generative architecture built from pairwise squeezing, local phase rotations, and a programmable interferometer, read out by homodyne detection of the position quadratures. In our model squeezing creates correlations, phases set their sign and magnitude, and the interferometer redistributes them across mode pairs, such the effect of each layer on the output correlation matrix becomes more explainable. We train the circuit classically on an empirical correlation matrix using only the covariance of the simulated state, and obtain continuous samples by simulated homodyne detection of the trained circuit. To reach beyond the Gaussian family we include a Kerr nonlinearity \cite{LloydBraunstein1999,Walschaers2021NonGaussian} and represent its weak-strength action as a coherent sum of Gaussian branches, which keeps training within Gaussian phase-space methods. We test the architecture on daily financial returns \cite{Cont2001,Dechant2026FinancialQGM} with up to twenty-mode circuits, reproducing an empirical correlation matrix up to small residuals and recovering the sign of correlations for most of the pairs.

\section{Model}
\label{sec:model}

We build a photonic circuit whose homodyne samples follow a multivariate distribution with tunable correlations. We use Gaussian gates with known effects on the position-quadrature correlations, add local Kerr gates to introduce non-Gaussianity, and train the circuit classically to match a target correlation matrix.

\subsection{Photonic variables and structured ansatz}
\label{sec:ansatz}

Each optical mode $j=1,\dots,N$ carries position and momentum quadratures $\hat{x}_j$ and $\hat{p}_j$ with commutation relations $[\hat{x}_j,\hat{p}_k]=2\,i\delta_{jk}$. We collect the quadratures into a vector $\hat{\mathbf{R}}=(\hat{x}_1,\hat{p}_1,\dots,\hat{x}_N,\hat{p}_N)^{T}$. A Gaussian state is specified by the first moments $\mathbf{d}=\langle\hat{\mathbf{R}}\rangle$ and the $2N\times2N$ covariance matrix
\begin{equation}
    \Sigma_{ab}=\langle\{\hat{R}_a-d_a,\hat{R}_b-d_b\}\rangle ,
    \label{eq:cov}
\end{equation}
with $\{\cdot,\cdot\}$ the anticommutator, and Gaussian gates act on $\Sigma$ as symplectic maps \cite{Weedbrook2012CVReview} (App.~\ref{app:background}). The vacuum then has $\Sigma / 2=\mathds{1}$, and Pearson correlations do not depend on this choice of normalization.

The model outputs are obtained by homodyne detection of $\hat{x}_j$ on every mode. The outcome $\bm{x}\in\mathbb{R}^N$ is distributed according to the position marginal of the Wigner function, which for a Gaussian state is a multivariate normal distribution with covariance given by the $N\times N$ position block $\Sx_{ij}=\Sigma_{2i-1,2j-1}$. The output correlations are then determined by
\begin{equation}
    (\Cgen)_{ij}=\Sx_{ij}/\sqrt{\Sx_{ii}\Sx_{jj}}.
    \label{eq:pearson}
\end{equation}
Note that our choice of measurement is distinct from boson sampling models, where photon-number-resolving detection of the same state returns integer counts with probabilities governed by hafnians \cite{Hamilton2017GBS}.
\begin{figure}[t]
\centering
\includegraphics[width=\columnwidth]{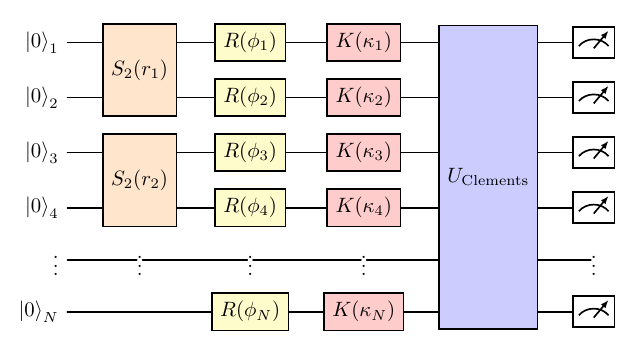}
\caption{$N$-mode CV photonic circuit. The modes start in the vacuum (left). Two-mode squeezers $S_2(r_j)$ act on the pairs $(1,2)$, $(3,4)$, $\dots$. For odd $N$ the unpaired last mode receives a single-mode squeezer instead. Every mode then passes through a local phase rotation $R(\phi_i)$ and a Kerr gate $K(\kappa_i)$, after which a Clements mesh $U_{\mathrm{Clements}}$ of $N(N-1)/2$ beam splitters mixes all modes. The output is read out by position-quadrature homodyne detection of every mode (right), which returns one real number per mode and hence one $N$-dimensional sample per shot. We treat squeezing amplitudes, local phases, and beam-splitter angles as trainable parameters, while keeping the Kerr gate parameter $\kappa$ fixed.}
\label{fig:ansatz}
\end{figure}

We prepare a CV photonic state by applying an ansatz $U(\bm{\theta})$ to the vacuum $\ket{\bm{0}}=\ket{0}^{\otimes N}$, where
\begin{equation}
    U(\bm{\theta})=U_{\mathrm{Clements}}(\bm{\vartheta})
    \prod_{i=1}^{N}K_i(\kappa)R_i(\phi_i)
    \prod_{(j,j')}S_{2}^{(j,j')}(r_{jj'}),
    \label{eq:ansatz}
\end{equation}
$(j,j')$ runs over the pairs $(1,2),(3,4),\dots$ and, for odd $N$, the last mode receives a single-mode squeezer $S(r_s)$ (Fig.~\ref{fig:ansatz}). The trainable parameters $\bm{\theta}$ are the squeezing amplitudes, the local phases $\phi_i$, and the angles $\bm{\vartheta}$ of the $N(N-1)/2$ beam splitters of the Clements mesh \cite{Clements2016Clements}. The internal phases of the beam splitters are fixed at zero, and the mesh represents a restricted real-orthogonal mixing layer. The homodyne measurement follows the Born rule and reads $p(\bm{x})=|\langle \bm{x}|U(\bm{\theta})|\bm{0}\rangle|^2$. In each run we then get samples $\bm{x} = (x_1, x_2, \dots, x_N) \sim p(\bm{x})$.

The three Gaussian layers shape the correlations in distinct ways. Two-mode squeezing correlates the position quadratures of a vacuum pair, with the magnitude set by the squeezing amplitude $r$. A local phase rotation changes which quadrature is measured, which tunes the sign and magnitude of the correlation at fixed $r$. The Clements mesh then redistributes these correlations across all modes (Sec.~\ref{sec:two-mode}). While the ansatz is simple, it spans a variety of correlated distributions and returns one $N$-dimensional sample per shot, with a potentially short time per sample.

Once the model is defined, we test it \textit{in silico} with established simulators of photonic systems, Strawberry Fields \cite{Killoran2019StrawberryFields} and Piquasso \cite{Kolarovszki2024Piquasso}. The simulation workflow splits into a Gaussian backend based on symplectic propagation, which is fast, and a Fock backend that works in a truncated Fock basis, which is numerically heavy and depends on the cutoff. We use Piquasso for performance, with the Gaussian backend for training and the Fock backend for testing the effects of the Kerr nonlinearity.

\subsection{Kerr nonlinearity and Gaussian-branch representation}
\label{sec:kerr}

Gaussian circuits with Gaussian measurements are efficiently simulable classically \cite{Bartlett2002Efficient,MariEisert2012}, and the simulation cost increases once non-Gaussian gates, such as a Kerr gate, enter the preparation circuit. These gates lead to output distributions outside the normal family and increase the expressivity of the generative model. We use the single-mode Kerr gate $K(\kappa)=\exp(i\kappa\hat{n}^2)$ with $\hat{n}=\hat{a}^\dagger\hat{a}$ \cite{Milburn1989KerrQC,Yurke1986}, whose quartic generator imprints an intensity-dependent phase and distorts the Wigner function.

To keep the simulation within Gaussian methods we represent the Kerr gate as a linear combination of Gaussian operations. This follows from the Hubbard--Stratonovich identity \cite{Hubbard1959,Stratonovich1957}, which writes $\exp(i\kappa\hat n^2)$ as an integral over $\exp(ix\hat n)$ with a Gaussian kernel,
\begin{equation}
    \exp(i\kappa\hat{n}^2)=\sqrt{\frac{i}{4\pi\kappa}}\int_{-\infty}^{\infty}
    e^{-i x^2/4\kappa}\,e^{i x\hat{n}}\,dx ,
    \label{eq:hs}
\end{equation}
for $\kappa>0$ with the principal branch of the square root. A quadrature rule for the oscillatory integrand \cite{Levin1996Oscillatory} gives a finite sum (App.~\ref{app:kerr}), and the action on a Gaussian state $\ket{G}$ becomes
\begin{equation}
    K(\kappa)\ket{G}\approx\sum_{m=1}^{M}c_m\ket{G_m},\qquad \ket{G_m}=R(x_m)\ket{G},
    \label{eq:branches}
\end{equation}
where the complex weights $c_m$ and real nodes $x_m$ come from the discretization of Eq.~\eqref{eq:hs}. Each branch $\ket{G_m}$ is Gaussian and propagates through the remaining layers within the Gaussian formalism. The branches add as amplitudes, so expectation values contain interference terms between every pair of branches,
\begin{equation}
\langle\hat{O}\rangle=\frac{\sum_{m,m'}c_m^{*}c_{m'}\braket{G_m|\hat{O}|G_{m'}}}{\sum_{m,m'}c_m^{*}c_{m'}\braket{G_m|G_{m'}}},
    \label{eq:branch-expectation}
\end{equation}
where the denominator normalizes the truncated sum. Overlaps and matrix elements of quadrature monomials between Gaussian states are available in closed form, so the correlators needed for Eq.~\eqref{eq:pearson} follow by choosing $\hat O=\hat x_i\hat x_j$ and $\hat O=\hat x_i$ and subtracting the means. The accuracy of the representation is controlled by the approximation parameters, in particular the number of nodes $M$, and the number of branches multiplies across modes. More details on this decomposition are provided in App.~\ref{app:kerr}.

In the following, we work in the regime of weak Kerr nonlinearity, with $\kappa=10^{-5}$ in the five-mode circuit and vanishing nonlinearity in the twenty-mode circuit, where the representation of Eq.~\eqref{eq:branch-expectation} requires a few thousand nodes per gate (App.~\ref{app:kerr}).

\subsection{Training, sampling, and evaluation}
\label{sec:training}

We train the model on an empirical dataset $\mathcal{D}_{\mathrm{target}}$ by fixing its output correlations to the target computed from data. Inference then returns real-valued multivariate samples with these correlations. From $T$ joint observations $\{\bm{x}^{(t)}\}_{t=1}^{T}$ of $N$ variables we compute the sample Pearson correlation matrix $\Ctgt$ and assign each variable to one mode. For a parameter vector $\bm{\theta}$ we evaluate the output state, extract $\Sx(\bm{\theta})$, and form $\Cgen(\bm{\theta})$ by Eq.~\eqref{eq:pearson}. The loss is a distance between the model and empirical matrices, measured as a maximum mean discrepancy (MMD) \cite{Gretton2012MMD}. With $\mathbf{r}_i$ the $i$-th row of a correlation matrix and the kernel $k(\mathbf{u},\mathbf{v})=\exp(-\gamma\|\mathbf{u}-\mathbf{v}\|^2)$ with $\gamma=1$, the MMD loss reads
\begin{equation}
\begin{split}
    \mathcal{L}_{\mathrm{MMD}}(\bm{\theta}) =
    \frac{1}{N^2}\sum_{i,j}\Big[& k(\mathbf{r}^{\mathrm{gen}}_i,\mathbf{r}^{\mathrm{gen}}_j)
    + k(\mathbf{r}^{\mathrm{target}}_i,\mathbf{r}^{\mathrm{target}}_j) \\
    & - 2\,k(\mathbf{r}^{\mathrm{gen}}_i,\mathbf{r}^{\mathrm{target}}_j)\Big].
\end{split}
\label{eq:mmd}
\end{equation}
We minimize Eq.~\eqref{eq:mmd} with the bound-constrained quasi-Newton method L-BFGS-B \cite{Byrd1995LBFGSB} under the box constraints of App.~\ref{app:training}. Training uses only the covariance of the simulated state and requires no sampling, so it corresponds to moment matching. Minimizing the distance between correlation matrices does not guarantee a match of the full model and empirical distributions, e.g. in total variation distance, but it reproduces the correlations and serves as a pre-training route. The marginals are adjusted in a post-processing stage applied to the generated samples.

\textit{Post-processing with affine maps.} After training we draw position-quadrature homodyne samples $\bm{x} := \bm{x}_{\mathrm{raw}}$ from the optimized circuit. The correlation loss is invariant under a rescaling of each mode, so the raw samples carry the learned correlations but arbitrary scales. We apply a per-mode affine map
\begin{equation}
    x_{\mathrm{final},i}=a_i\,x_{\mathrm{raw},i}+b_i ,
    \label{eq:affine}
\end{equation}
with $a_i>0$ and $b_i$ chosen to match the mean and variance of each generated marginal to the target. For positive $a_i$ the Pearson correlations are unchanged. The map adjusts location and scale and cannot alter the shape of a marginal.

We evaluate the models with quantities distinct from the training loss. For correlation agreement we use the mean squared error $\mathrm{MSE}_{\mathrm{full}}=N^{-2}\sum_{i,j}[(\Ctgt)_{ij}-(\Cgen)_{ij}]^2$ over all entries, the entrywise residuals $\Delta C_{ij}=(\Ctgt)_{ij}-(\Cgen)_{ij}$, and the sign-agreement rate, the fraction of the $N(N-1)/2$ pairs $i<j$ for which $\Cgen$ and $\Ctgt$ have the same sign. For the marginals we use the total variation distance $\mathrm{TVD}=\tfrac12\sum_b|p_b-q_b|$ between histograms of generated and target samples and the two-sample Kolmogorov--Smirnov (KS) statistic $D$ with its $p$-value. Estimator settings are given in App.~\ref{app:training}.
\begin{figure}[t]
    \centering
    \includegraphics[width=\columnwidth]{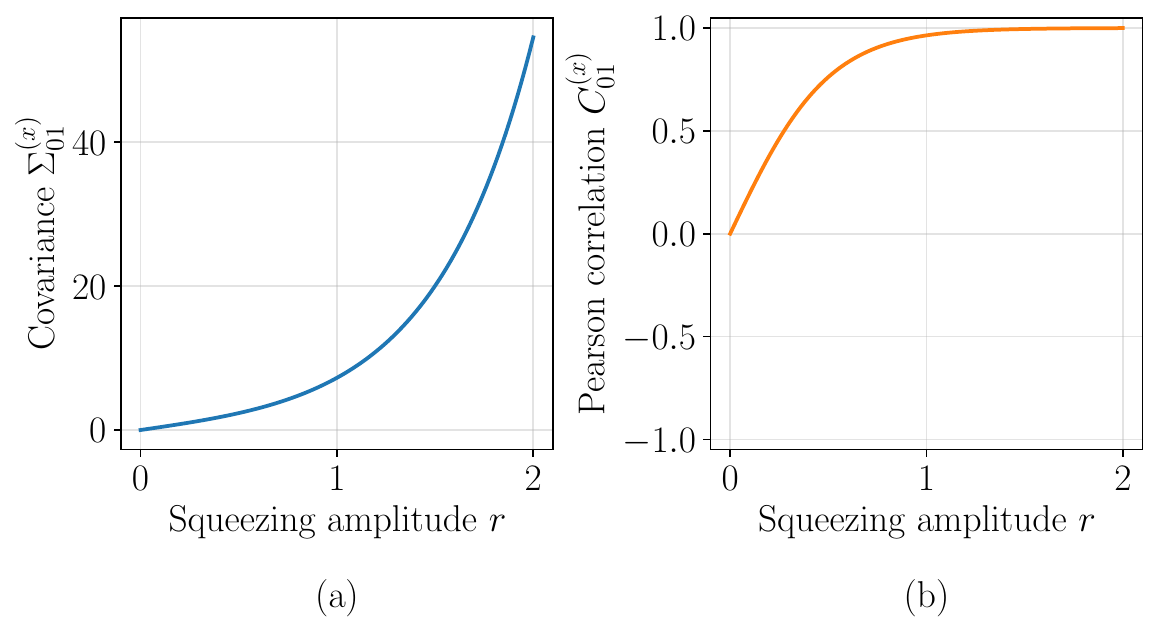}
    \caption{Correlation control by two-mode squeezing. A two-mode squeezed vacuum $S_2(r)\ket{0,0}$ is simulated deterministically for squeezing amplitudes $r\in[0,2]$. (a)~Position-quadrature covariance between the two modes, which grows exponentially with $r$. (b)~Pearson correlation between the two position quadratures, which rises steeply at small $r$, exceeds $0.9$ for $r\gtrsim0.8$, and saturates toward unity without reaching it at finite $r$. The squeezing layer therefore sets the magnitude of a pairwise correlation. The axis labels index the two modes as 0 and 1.}
    \label{fig:two-mode}
\end{figure}

\section{Results}
\label{sec:results}

We now test the approach, which requires continuous multivariate datasets with a nontrivial correlation structure. We choose financial datasets, which exhibit positive and negative correlations, multivariate trends, and heavy tails, while retaining a significant normal component \cite{Cont2001,McNeil2005QRM}. Specifically, we collect daily returns of several assets over a fixed observation window. Daily adjusted closing prices $P_t$ are retrieved with \texttt{yfinance} \cite{Aroussi2023yfinance}, converted to log returns $r_t=\log P_t-\log P_{t-1}$, and standardized to zero mean and unit variance per asset. The target $\Ctgt$ is the sample Pearson correlation matrix over the window, and we study joint return statistics at a fixed window without any forecasting. Two ensembles are used: a five-asset set consisting of the equities AAPL and MSFT, the gold fund GLD, the long-maturity Treasury fund TLT, and the bitcoin-dollar pair BTC-USD over $T=251$ daily returns, and a set of twenty S\&P~500 equities over $T=48$ daily returns. We present more details on acquisition and observation windows in App.~\ref{app:training}.

\subsection{Two-mode mechanism}
\label{sec:two-mode}

Before training in the general $N$-mode case, we examine the properties of the model on a two-mode example. In Fig.~\ref{fig:two-mode} we show the position-quadrature covariance and the Pearson correlation of a two-mode squeezed vacuum for $r\in[0,2]$. The covariance grows exponentially with $r$, while the correlation rises steeply at small $r$, exceeds $0.9$ for $r\gtrsim0.8$, and saturates toward unity without reaching it at finite squeezing. Squeezing therefore creates correlation between a mode pair and sets its magnitude. With the squeezing fixed, a phase rotation applied to one mode of the pair and swept through one period modulates the measured correlation as a cosine of the applied phase, from its maximal positive value through zero at $\pi/2$ to its maximal negative value at $\pi$ (App.~\ref{app:extra}). A local phase thus sets the sign of a measured correlation and tunes its magnitude continuously. The mesh then mixes the modes and redistributes these correlations across all pairs.
\begin{figure*}[t]
    \centering
    \includegraphics[width=0.98\textwidth]{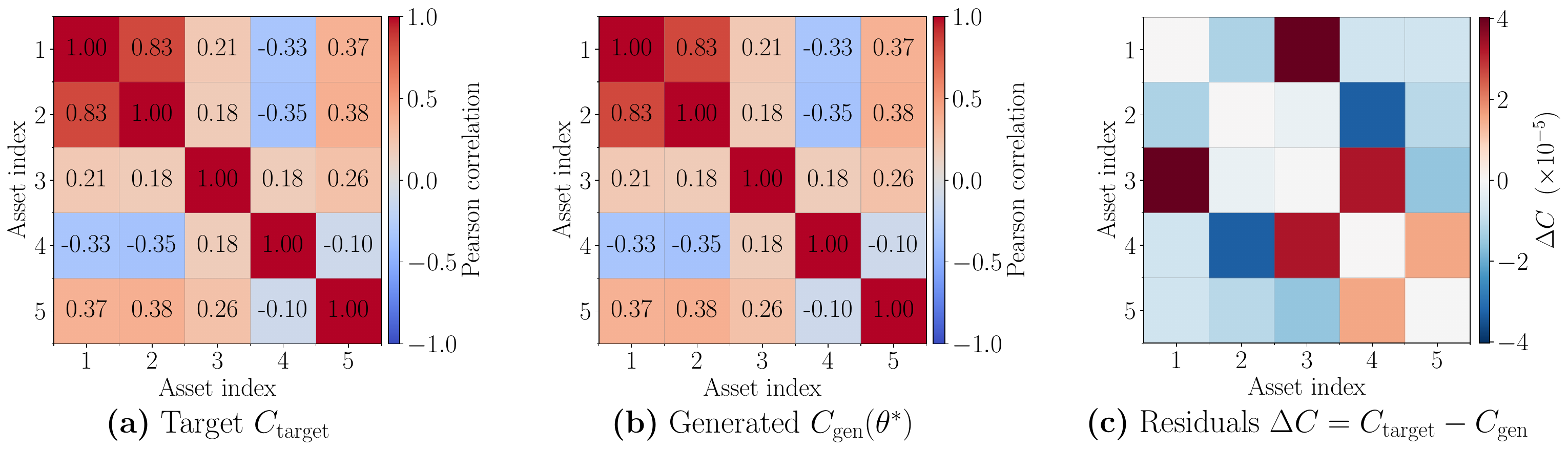}
    \caption{Correlation learning with the five-mode circuit on the five-asset dataset. Asset indices 1 to 5 denote AAPL, MSFT, GLD, TLT, and BTC-USD. (a)~Empirical Pearson correlation matrix $\Ctgt$ of the standardized daily log returns over $T=251$ days, the training target. (b)~Correlation matrix $\Cgen(\bm{\theta}^{*})$ of the position-quadrature output of the optimized circuit. (c)~Entrywise residuals $\Delta C_{ij}=(\Ctgt)_{ij}-(\Cgen)_{ij}$ on a color scale of $\pm 4.1\times10^{-5}$.}
    \label{fig:five-corr}
\end{figure*}

Using this interpretation, the model becomes a design tool for distributions in which positive and negative correlations are defining properties. For instance, an inspection of an empirical correlation matrix shows which pairs are strongly correlated and with which sign, which suggests a suitable initialization. Based on this, we select and pair modes for two-mode squeezing, choose the initial squeezing amplitudes, and decide which local phases start near $0$ or $\pi$. Such a physically informed initialization can precede numerical optimization or serve as a pre-training stage before the mesh parameters are refined.

\subsection{Five-mode demonstration}
\label{sec:five-mode}

\begin{figure}[b]
    \centering
    \includegraphics[width=\columnwidth]{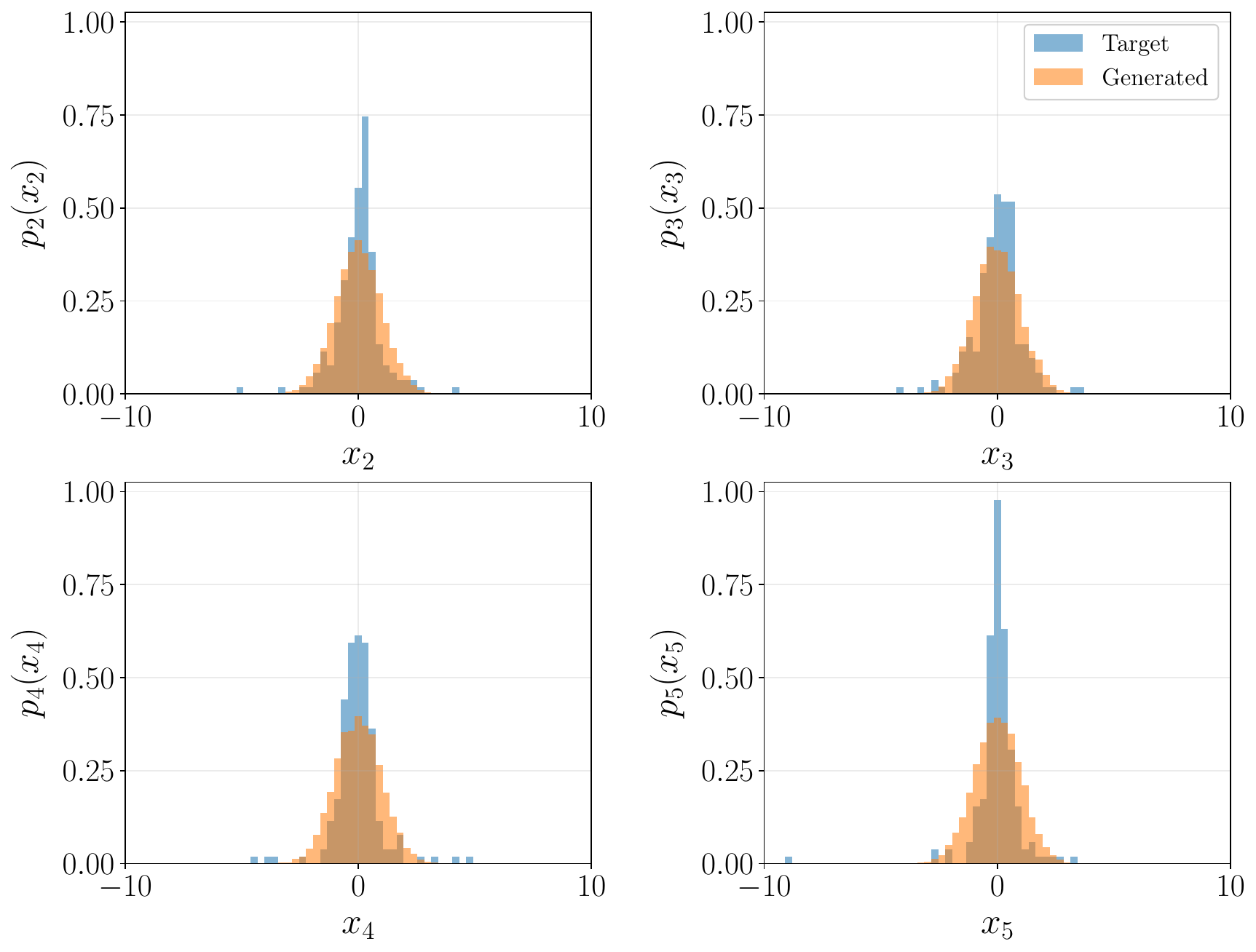}
    \caption{Univariate marginals of the five-mode model after the per-mode affine map [Eq.~\eqref{eq:affine}], for assets 2 to 5 (MSFT, GLD, TLT, and BTC-USD; AAPL is omitted for brevity, see Table~\ref{tab:marginals}). Blue histograms: standardized target daily log returns. Orange histograms: position-quadrature homodyne samples from the optimized circuit, with mean and variance matched to the target.}
    \label{fig:five-marg}
\end{figure}

We proceed to model a multivariate distribution from the financial domain, instantiating the ansatz with $N=5$. Two-mode squeezers are applied on pairs $(1,2)$ and $(3,4)$, and a single-mode squeezer on mode~5. Local phase rotations and single-mode Kerr gates with $\kappa=10^{-5}$ act on every mode, followed by a Clements mesh of ten beam splitters. The circuit has $18$ trainable parameters, corresponding to five phases, three squeezing amplitudes, and ten beam-splitter angles. We minimize Eq.~\eqref{eq:mmd} with L-BFGS-B from one fixed initialization, which is the run shown in Figs.~\ref{fig:five-corr} and \ref{fig:five-marg}, as well as from ten random perturbations around that initialization (App.~\ref{app:training}).

We compare the target and generated matrices in Fig.~\ref{fig:five-corr}. The optimized circuit reproduces the sign and magnitude of all ten pairwise correlations, including the negative correlations of TLT with the two equities and the weak correlations involving GLD. The residuals are bounded within $4.1\times10^{-5}$, with $\mathrm{MSE}_{\mathrm{full}}=3.7\times10^{-10}$. All ten perturbed initializations reach $\mathrm{MSE}_{\mathrm{full}}$ below $3.1\times10^{-8}$ within 4600 function evaluations (App.~\ref{app:extra}).

Fig.~\ref{fig:five-marg} shows the post-affine homodyne samples beyond their second moments. The raw generated samples are unimodal and centered near zero, with widths that differ across assets because the loss does not constrain the scale of each mode (App.~\ref{app:extra}, Table~\ref{tab:marginals}). The affine map brings every generated marginal onto the target mean and variance and visibly improves the overlap in the central region. We note that the targets have sharper peaks and heavier tails than a normal distribution, while the marginals of the trained circuit, which is Gaussian up to the weak Kerr distortion, remain close to normal under any affine map. The mismatch is largest for the most volatile asset, BTC-USD. Quantitatively (App.~\ref{app:extra}, Table~\ref{tab:marginals}), the affine map reduces the asset-averaged TVD from $0.370$ to $0.245$ and the averaged KS statistic from $0.190$ to $0.121$, with BTC-USD retaining the largest TVD of $0.330$. For GLD, whose raw marginal already lies close to the target, the affine map raises the TVD from $0.165$ to $0.225$ and the KS statistic from $0.066$ to $0.087$, and for TLT it leaves the TVD essentially unchanged ($0.273$ and $0.271$). The correlation structure is thus learned to high accuracy. Learning the tails of the marginals and higher-order correlations requires further increasing the expressivity of the generative model with non-Gaussian gates, at the expense of more complex simulation.

\subsection{Twenty-mode demonstration}
\label{sec:twenty-mode}

\begin{figure*}[t]
    \centering
    \includegraphics[width=0.98\textwidth]{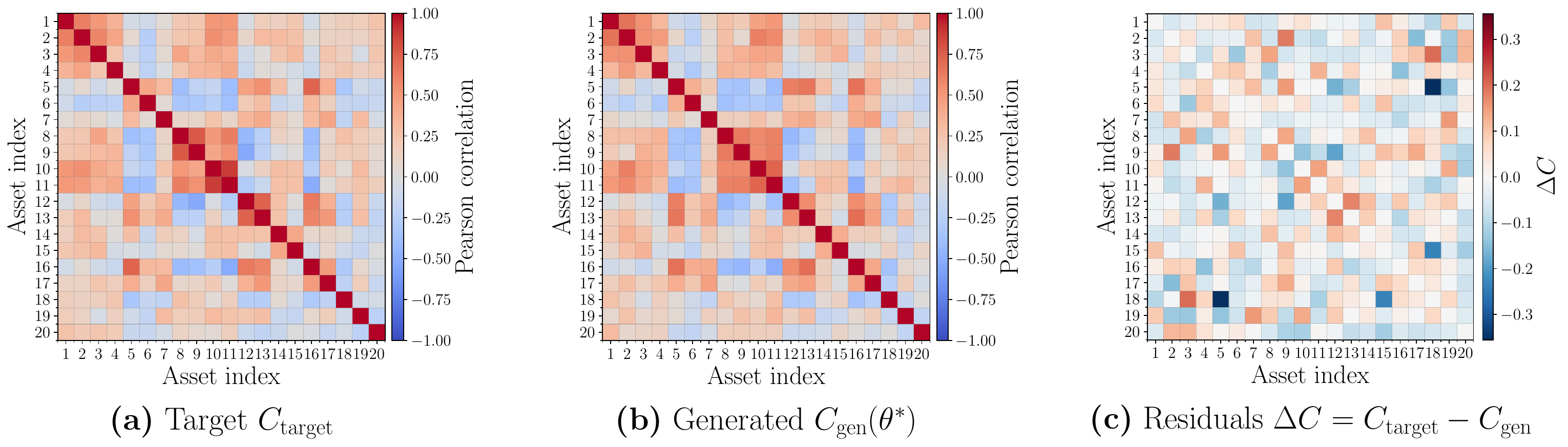}
    \caption{Correlation learning with a twenty-mode Gaussian circuit (220 trainable parameters) on twenty S\&P~500 equities. (a)~Empirical Pearson correlation matrix $\Ctgt$ of the standardized daily log returns over $T=48$ trading days. (b)~Correlation matrix $\Cgen(\bm{\theta}^{*})$ of the optimized circuit for the run started from the fixed initialization of App.~\ref{app:training}. (c)~Entrywise residuals $\Delta C_{ij}=(\Ctgt)_{ij}-(\Cgen)_{ij}$ on a color scale of $\pm0.36$. The block of strongly correlated equities around assets 8 to 12 and the weaker, partly negative correlations elsewhere are reproduced. Aggregate performance over five random initializations (sign agreement and residual distribution) is reported in the text and in App.~\ref{app:extra}, Table~\ref{tab:residuals}.}
    \label{fig:twenty-corr}
\end{figure*}

We extend our tests to $N=20$ multivariate distributions without structural changes. The model includes ten two-mode squeezers on consecutive pairs, a local phase on every mode, and a mesh of 190 beam splitters. With zero Kerr nonlinearity this is a Gaussian circuit with 220 trainable parameters (20 phases, 10 squeezing amplitudes, 190 beam-splitter angles), trained on the empirical correlation matrix of twenty equities from a fixed initialization, which is the run shown in Fig.~\ref{fig:twenty-corr}, and from its five random perturbations.

In Fig.~\ref{fig:twenty-corr} we compare the matrices for one optimized circuit. The generated matrix reproduces the block of strongly correlated equities around modes 8 to 12 and the weaker, partly negative correlations elsewhere. Averaged over seeds, the sign of $93.8\%\pm0.9\%$ of the 190 pairs is recovered (standard deviation across seeds, individual runs from $92.6\%$ to $94.7\%$), $86\%$ of the pairs have residuals below $0.10$, and no residual exceeds $0.5$ (App.~\ref{app:extra}, Table~\ref{tab:residuals}). Since equity correlations are predominantly positive, a constant positive sign would also score well above $50\%$, so we read the sign agreement together with the residual distribution. All five runs reach a comparable plateau of $\mathrm{MSE}_{\mathrm{full}}$ (App.~\ref{app:extra}, Fig.~\ref{fig:twenty-conv}). The target is estimated from 48 observations of 20 variables and is a noisy estimate of the dependence between the assets, and the model is trained to reproduce this empirical matrix. The experiment shows that the ansatz and the correlation loss remain trainable with 220 parameters and 190 target entries at this size.

\section{Discussion}
\label{sec:discussion}

We first place the model relative to boson sampling. Gaussian boson sampling \cite{Lund2014GaussianBS,Hamilton2017GBS}, its Fock-state and nonlinear variants \cite{Aaronson2011BosonSampling,Spagnolo2023NonlinearBS}, and the advantage experiments built on it \cite{Zhong2020Jiuzhang,Madsen2022PhotonicAdvantage} use squeezers and an interferometer (as in Fig.~\ref{fig:ansatz}) but read the output with photon-number-resolving detectors. Their hardness is based on the readout. We read the same states with homodyne detection, which gives a Gaussian output whose covariance we compute and sample classically in polynomial time \cite{Bartlett2002Efficient,MariEisert2012}. This makes the training classically tractable, while hardness results for homodyne-sampled circuits require non-Gaussian elements of finite strength \cite{Douce2017CVIQP,Chabaud2017CVSampling,Chabaud2023Resources}. To gain a computational advantage one therefore needs models with larger $\kappa$, at the expense of simulation complexity.

Next, we discuss experimental benefits and limitations of CV optics. Continuous-variable systems offer a versatile optical platform \cite{Andersen2015HybridCV,Flamini2019PhotonicQML}, but photon loss is the main limitation for CV-based processing \cite{Weedbrook2012CVReview,Niset2009NoGo}. Consider a loss channel that admixes vacuum into each mode with transmission $\eta$. At $\eta = 1$ and for $8.7$~dB ($r=1$) of squeezing the correlation reaches $0.96$. At $\eta=0.5$ it drops to $0.76$, and recovering $0.96$ then requires about $17$~dB of squeezing. Bulk optics reaches $15$~dB \cite{Vahlbruch2016Squeezing15dB}, while integrated sources reach $1$~dB measured ($4$~dB inferred on chip) in silicon nitride \cite{Vaidya2020NanophotonicSqueezing} and $4.9$~dB measured ($11$~dB inferred) in thin-film lithium niobate \cite{Nehra2022FewCycleSqueezing}, which corresponds to end-to-end transmissions of roughly $0.3$ to $0.7$. As Gaussian error correction cannot remove this constraint \cite{Niset2009NoGo}, the model favors a shallow circuit architecture, possibly with a modified interferometric mesh.

At the same time, optical systems offer short times per sample at large width. Homodyne receivers run at gigahertz bandwidth \cite{Tasker2021GHzHomodyne}, multimode homodyne detection reaches $60$ modes \cite{Yoon2026Homodyne}, and every shot returns a full $N$-dimensional sample. For a Gaussian circuit a classical sampler competes at $O(N^2)$ per sample, and larger width does not provide a large separation on its own. With the addition of the Kerr layer this changes. The branch count (Sec.~\ref{sec:kerr}) multiplies across modes, and phase-space expansions for stronger non-Gaussianity carry a cost that grows with the non-Gaussian content \cite{Bourassa2021,Guseynov2026CoherentStatePropagation,Walschaers2021NonGaussian}, whereas the physical sampling cost does not depend on $\kappa$. We regard a wide, shallow circuit with Kerr gates strong enough to make the branch sum expensive, benchmarked by time per sample at fixed correlation accuracy, as the target. The Gaussian boson sampler of Ref.~\cite{Cimini2026GBSReservoir}, with more than $400$ frequency modes used as a reservoir in which inter-mode correlations carried the learned information, shows that Gaussian optics at this width is available and that its correlations are useful for learning.

\section{Conclusion}
\label{sec:conclusion}

We proposed a photonic generative model based on continuous-variable encoding and a readout of position quadratures by homodyne detection. The model is built from pairwise squeezing, local phase adjustment, and interferometric mixing, where each layer controls the correlations in an interpretable way. We trained it by minimizing the distance between empirical correlation matrices and those generated by the model, reproducing correlations between multiple stochastic variables corresponding to log returns of financial assets. We also suggested an effective simulation method for a weak Kerr gate based on a coherent sum of Gaussian branches, which keeps the whole pipeline within symplectic simulation methods. Given that fast optical measurements provide real-valued samples from a joint multivariate distribution, this generative model represents a native sampler for certain distribution families at a small time-per-sample cost. As the next step, we consider models with significant non-Gaussian content, via stronger Kerr gates or measurement-induced nonlinearities, and the transfer of trained circuits onto photonic hardware with realistic loss.

\begin{acknowledgments}
We thank Zolt\'an Kolarovszki and Zolt\'an Zimbor\'as for insightful discussions on simulation backends. We are grateful to Raj B. Patel and Shang Yu for the discussions of experimental realizations. We also thank Yuan Wang for helping with the submission, and Pieter Kok for useful suggestions. O.~K. acknowledges the support from the UK EPSRC award for QCi3 Hub (EP/Z53318X/1) and Open Plus Fellowship award ``Quantum Generative Modelling: from Foundations to Applications''.
\end{acknowledgments}

\vspace{10pt}
\textit{Contributions and AI statement.} The original idea was proposed by O.\,K. and developed by D.\,C.\,L. as a part of the MSc project starter over a year ago, before agents became widespread. When finalizing the project, we used AI tools for writing parts of the manuscript and polishing the code. The authors take responsibility for results and presentation.

\appendix

\section{Gaussian states and optical gates}
\label{app:background}

In this appendix we collect the phase-space conventions behind Eqs.~\eqref{eq:cov} and \eqref{eq:pearson}, and summarize the action of the gates that enter the ansatz of Eq.~\eqref{eq:ansatz}. This appendix supports Sec.~\ref{sec:ansatz}.

The state of $N$ optical modes lives in $\mathcal{H}=\bigotimes_{j=1}^{N}\mathcal{H}_j$. A state is Gaussian when its Wigner function is a Gaussian function of the phase-space variables, in which case the first moments $\mathbf{d}$ and the covariance matrix of Eq.~\eqref{eq:cov} determine it completely \cite{Braunstein2005CVReview,Weedbrook2012CVReview}. With the commutator $[\hat x,\hat p]=2i$ (Sec.~\ref{sec:ansatz}) the vacuum covariance matrix is $\Sigma/2=\mathds{1}$, and a two-mode squeezed vacuum has $\Sigma^{(x)}_{12}=2\sinh 2r$, the curve in Fig.~\ref{fig:two-mode}(a). Measuring a subset of modes yields outcomes distributed according to the corresponding Wigner marginal, and the position marginal of any state is the density of position-quadrature homodyne outcomes. Gaussian unitaries are generated by Hamiltonians at most quadratic in the quadratures and act as $\mathbf{d}\mapsto S\mathbf{d}$ and $\Sigma\mapsto S\Sigma S^{T}$ with $S$ symplectic.

In Table~\ref{tab:gates} we list the gates of Eq.~\eqref{eq:ansatz} together with the rule that each one imposes on the measured correlations. The first four are Gaussian and account for the correlation mechanism of Sec.~\ref{sec:two-mode}. The Kerr gate is the single non-Gaussian element. Its generator $\hat n^2$, with $\hat{n}=\tfrac{1}{4}(\hat{x}^2+\hat{p}^2)-\tfrac{1}{2}$, is quartic in the quadratures and has no symplectic representation. In general this breaks the Gaussian structure of the model, and in Sec.~\ref{sec:kerr} we represent it as a sum of Gaussian operations.
\begin{table}[h]
\centering
\footnotesize
\setlength{\tabcolsep}{3pt}
\begin{tabular}{@{}p{0.2\columnwidth}p{0.1\columnwidth}p{0.63\columnwidth}@{}}
\hline\hline
Operation & Symbol & Rule for the measured correlations \\
\hline
Two-mode squeezing & $S_2(r)$ & Creates a positive correlation between the position quadratures of a pair; the magnitude grows monotonically with $r$ and saturates toward unity [Fig.~\ref{fig:two-mode}(b)]. \\
Single-mode squeezing & $S(r_s)$ & Rescales the quadratures of an unpaired mode; its correlations with other modes arise only after mixing in the mesh. \\
Phase rotation & $R(\phi)$ & Multiplies the correlation of a squeezed pair by $\cos\phi$: $\phi=0$ keeps it, $\phi=\pi/2$ removes it, $\phi=\pi$ flips its sign (Fig.~\ref{fig:phase}). \\
Beam splitter & $B(\vartheta)$ & Mixes two modes; the Clements mesh redistributes the pair correlations across all modes, with $\vartheta=\pi/4$ a balanced split. \\
Kerr gate & $K(\kappa)$ & Intensity-dependent phase $e^{i\kappa\hat{n}^2}$; adds non-Gaussian shape to the output, while at small $\kappa$ the second moments are nearly unchanged. \\
\hline\hline
\end{tabular}
\caption{Optical gates of the ansatz in Eq.~\eqref{eq:ansatz} and the rule by which each one controls the correlations measured by position-quadrature homodyne detection.}
\label{tab:gates}
\end{table}

\section{Kerr decomposition: construction, settings, and checks}
\label{app:kerr}

In this appendix we describe the integral representation of a single-mode Kerr gate and the finite-sum construction behind the branch expansion of Eq.~\eqref{eq:branches}, and we report the errors of the decomposition against exact Kerr evolution on Fock states. The scope is limited to the weak-nonlinearity regime used in Sec.~\ref{sec:five-mode}. A systematic study of the representation is the subject of forthcoming work.

\emph{Integral representation.} The Hubbard-Stratonovich identity of Eq.~\eqref{eq:hs} linearizes the quadratic exponent of the Kerr gate at the cost of an auxiliary integral over $x$. Each $e^{ix\hat n}$ is a phase rotation $R(x)$, which is the origin of the Gaussian branches in Eq.~\eqref{eq:branches}. Choosing sufficiently large integration limits and a quadrature rule, we convert the integral into a sum of Gaussian operations.

\emph{Finite-sum construction.} The integrand oscillates rapidly, and to mitigate this we discretize the integral with Levin's collocation method \cite{Levin1996Oscillatory}, which converts the oscillatory integral into a small linear system and yields the finite sum $e^{i\kappa\hat n^2}\approx\sum_{m=1}^{M}c_m R(x_m)$ with complex weights $c_m$ and real nodes $x_m$. Applied to a Gaussian state this sum produces the branches $\ket{G_m}=R(x_m)\ket{G}$ of Eq.~\eqref{eq:branches}, and expectation values follow from Eq.~\eqref{eq:branch-expectation}. We set the limits to $[-100,100]$ and introduce a Gaussian window $e^{-\gamma_w x^2}$, with $\gamma_w = 5 \times 10^{-4}$, which regularizes the integrand.
\begin{figure}[t]
    \centering
    \includegraphics[width=0.92\columnwidth]{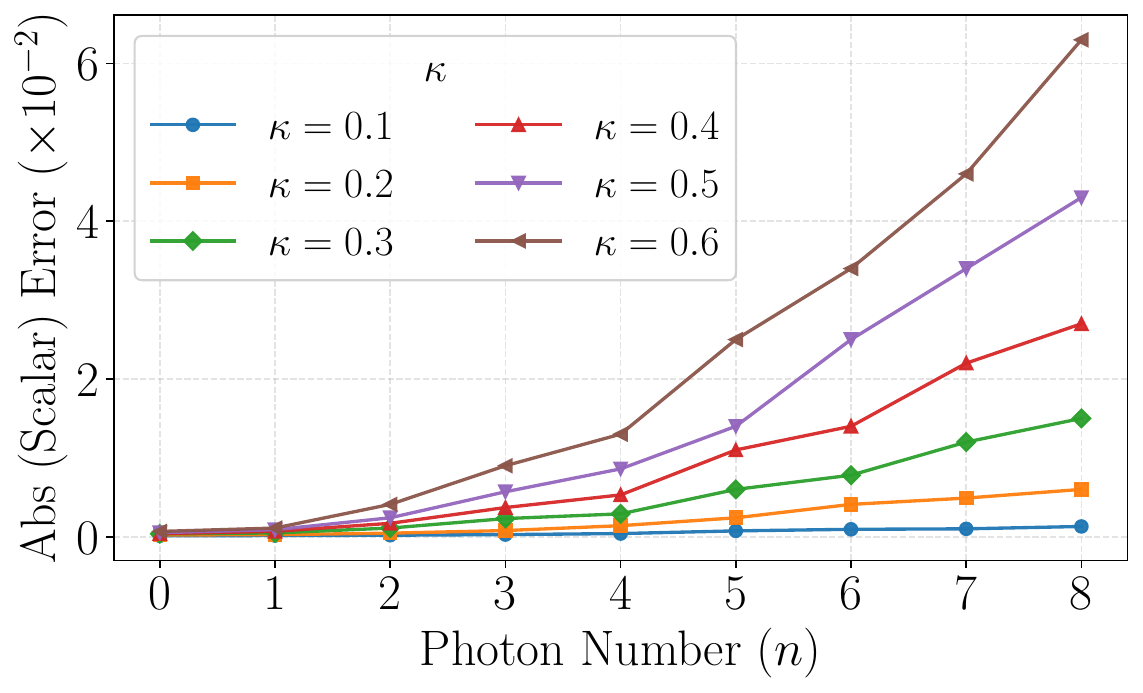}
    \caption{Error scaling of the finite-sum Kerr decomposition as a function of Fock occupation. Scalar error $\tilde\varepsilon_\kappa(n)$ between the exact Kerr phase and the decomposed sum (see text) for $n=0,\dots,8$ and six Kerr strengths $\kappa=0.1$ to $0.6$. The error increases with both $n$ and $\kappa$, stays below $1.3\times10^{-3}$ for $\kappa=0.1$ over the displayed range, and reaches about $6\times10^{-2}$ at $\kappa=0.6$ and $n=8$.}
    \label{fig:kerr-error}
\end{figure}

\emph{Approximation quality.} In Fig.~\ref{fig:kerr-error} we compare the decomposition with exact Kerr evolution on Fock states, where $K(\kappa)\ket{n}=e^{i\kappa n^2}\ket{n}$. As a reference we use the Kerr gate simulated in the Fock backend of Piquasso, and track the error of the sum $s_\kappa(n)=\sum_m c_m e^{ix_m n}$ against the exact phase. The plotted scalar error is
\begin{equation}
\begin{split}
    \tilde\varepsilon_\kappa(n)=\,&\big||\mathrm{Re}\,e^{i\kappa n^2}|-|\mathrm{Re}\,s_\kappa(n)|\big| \\
    &+\big||\mathrm{Im}\,e^{i\kappa n^2}|-|\mathrm{Im}\,s_\kappa(n)|\big| ,
\end{split}
    \label{eq:kerr-error}
\end{equation}
which for $n\ge1$ lies between 1.0 and 1.5 times the absolute error $|e^{i\kappa n^2}-s_\kappa(n)|$. The error grows with $n$ and with $\kappa$. At $\kappa=0.1$ it remains below $1.3\times10^{-3}$ over the displayed photon numbers, and at $\kappa=0.6$ it reaches about $6\times10^{-2}$ at $n=8$. The test establishes that the finite sum reproduces the Kerr phases accurately for weak nonlinearity and low occupation, which is the regime of the five-mode circuit. The operating value $\kappa=10^{-5}$ lies below the scanned range and was checked directly with the same Levin weights. The unnormalized sum carries a global factor of modulus about $2.2$, so $s_\kappa(n)$ differs from $e^{i\kappa n^2}$ by order one at this $\kappa$, but dividing by its value at $n=0$, which is the normalization of Eq.~\eqref{eq:branch-expectation}, gives an error below $3\times10^{-7}$ for $n\le8$.

\section{Datasets, training settings, and estimators}
\label{app:training}

In this appendix we list the details needed to reproduce the training experiments of Sec.~\ref{sec:results}: the construction of the two datasets and their correlation targets, the optimizer settings and parameter bounds used with the loss of Eq.~\eqref{eq:mmd}, and the conventions of the evaluation metrics defined in Sec.~\ref{sec:training}.

\emph{Datasets.} The five-asset dataset contains $T=251$ daily log returns of AAPL, MSFT, GLD, TLT, and BTC-USD. The twenty-equity dataset contains $T=48$ daily log returns of twenty S\&P~500 constituents. Prices are daily adjusted closes from Yahoo!~Finance via \texttt{yfinance} \cite{Aroussi2023yfinance}. Log returns are formed per asset, standardized to zero mean and unit variance over the window, and the target $\Ctgt$ is their sample Pearson correlation matrix. The five-asset window runs from 2020-03-02 to 2021-03-01, with 252 closing prices and $T=251$ returns. BTC-USD closes are sampled on the equity trading days, so its returns span weekends and holidays. The twenty-equity window runs from 2019-11-01 to 2020-01-13, with 49 closing prices and $T=48$ returns, and the tickers in mode order are AAPL, MSFT, GOOGL, NVDA, ED, NEM, VZ, XOM, CVX, JPM, BAC, PG, KO, JNJ, UNH, NEE, DUK, BA, WMT, and TSLA.

\emph{Training.} All circuits are trained with L-BFGS-B on Eq.~\eqref{eq:mmd} with kernel bandwidth $\gamma=1$ under box constraints that keep the simulated states in a numerically stable range: the local phases $\phi$ are unconstrained, the two-mode and single-mode squeezing amplitudes $r$ and $r_s$ lie in $[0,1.5]$, and the beam-splitter angles $\vartheta$ of the Clements mesh lie in $[0,\pi]$. The beam-splitter internal phases are fixed at zero, and the Kerr strength is fixed at $\kappa=10^{-5}$ in the five-mode circuit and $\kappa=0$ in the twenty-mode circuit. Each circuit is trained from a fixed initialization and from random perturbations of it. The fixed initialization sets all local phases to $0$, all beam-splitter angles to $\pi/4$, and all squeezing amplitudes to $0.35$ in the five-mode circuit and to $0.70$ in the twenty-mode circuit. It is the run shown in Figs.~\ref{fig:five-corr}, \ref{fig:five-marg}, and \ref{fig:twenty-corr}. The random initializations add to the phases and the beam-splitter angles uniform offsets in $[-0.3,0.3]$ and to the squeezing amplitudes uniform offsets in $[-0.1,0.1]$, with the amplitudes and angles clipped to their bounds. The five-mode circuit is trained from ten random initializations for up to about 4600 function evaluations, and the twenty-mode circuit from five random initializations for up to about 15000 function evaluations. The convergence plots count every function evaluation. The resulting convergence traces are shown in App.~\ref{app:extra}.

\emph{Estimators.} $\mathrm{MSE}_{\mathrm{full}}$ is averaged over all $N^2$ entries, including the diagonal, which contributes zero. TVD is computed between histograms of target and generated samples on a common binning of 50 equal-width bins spanning the combined range of both samples, and the KS test is the two-sample test on the same samples. The generated marginals use $10^4$ homodyne samples drawn from the normal distribution with covariance $\Sx$, and the target marginals use the first 70\% of the standardized returns (175 days, standardized with the mean and standard deviation of this training split).

\section{Additional results}
\label{app:extra}

In this appendix we collect results that support the findings in the main text: the phase control of the two-mode correlation (Sec.~\ref{sec:two-mode}), the marginal statistics of the five-mode model (Sec.~\ref{sec:five-mode}), the residual distribution of the twenty-mode model (Sec.~\ref{sec:twenty-mode}), and the convergence of the training runs at both sizes.

\begin{figure}[t]
    \centering
    \includegraphics[width=0.7\columnwidth]{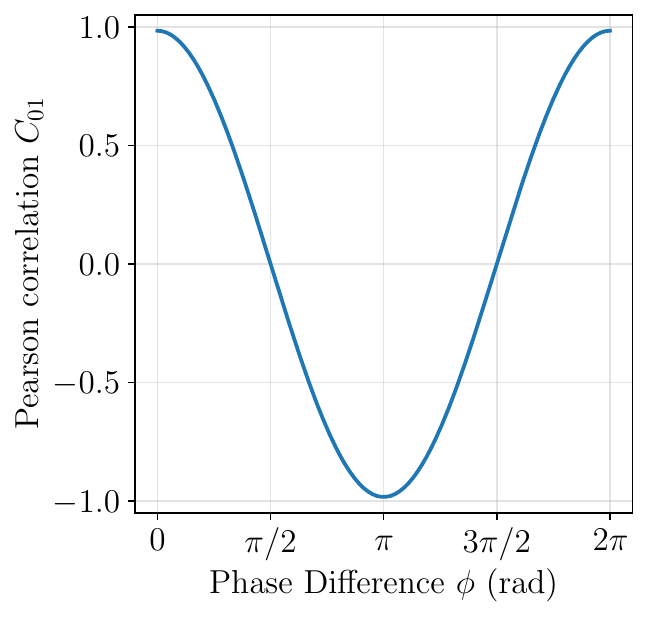}
    \caption{Phase control of the two-mode correlation. Pearson correlation between the position quadratures of a two-mode squeezed vacuum at fixed squeezing, as a function of a phase rotation $\phi$ applied to one of the two modes and swept over one period. The correlation follows a cosine of $\phi$ (maximal at $\phi=0$, crossing zero at $\phi=\pi/2$).}
    \label{fig:phase}
\end{figure}

\emph{Phase control.} Fig.~\ref{fig:phase} shows the second control discussed in Sec.~\ref{sec:two-mode}. For a two-mode squeezed vacuum at fixed squeezing, a phase rotation applied to one mode and swept over one period modulates the measured position-quadrature correlation as a cosine of the phase. Together with Fig.~\ref{fig:two-mode}, this shows that squeezing sets the magnitude of a pairwise correlation and a local phase sets its sign.

\emph{Five-mode marginal statistics.} Table~\ref{tab:marginals} quantifies the marginal comparison of Fig.~\ref{fig:five-marg} with the TVD and KS statistics of Sec.~\ref{sec:training}, before and after the affine map of Eq.~\eqref{eq:affine}. The affine map lowers the asset-averaged TVD from $0.370$ to $0.245$ and the averaged KS statistic from $0.190$ to $0.121$. The largest post-affine TVD, $0.330$, belongs to BTC-USD, the most volatile asset. For GLD the raw marginal has the smallest TVD ($0.165$) and a KS $p$-value of $0.43$, and the affine map raises its TVD to $0.225$ and lowers the $p$-value to $0.14$. For TLT the affine map changes neither the TVD ($0.273$ and $0.271$) nor, appreciably, the $p$-value ($0.0085$ and $0.0059$). The residual TVD across all assets reflects the heavy tails of the returns, which second-moment control does not reproduce.
\begin{table}[t]
    \centering
    \footnotesize
    \setlength{\tabcolsep}{3pt}
    \begin{tabular}{@{}lcccccc@{}}
        \hline\hline
        & \multicolumn{3}{c}{Before affine map} & \multicolumn{3}{c}{After affine map} \\
        \cline{2-4}\cline{5-7}
        Asset & TVD & KS $D$ & $p$-value & TVD & KS $D$ & $p$-value \\
        \hline
        AAPL    & 0.310 & 0.138 & 0.0026             & 0.202 & 0.090 & 0.112 \\
        MSFT    & 0.406 & 0.230 & $1.8\times10^{-8}$  & 0.195 & 0.102 & 0.051 \\
        GLD     & 0.165 & 0.066 & 0.431              & 0.225 & 0.087 & 0.138 \\
        TLT     & 0.273 & 0.125 & 0.0085             & 0.271 & 0.129 & 0.0059 \\
        BTC-USD & 0.696 & 0.390 & $5.3\times10^{-24}$ & 0.330 & 0.195 & $3.4\times10^{-6}$ \\
        \hline
        Average & 0.370 & 0.190 & --- & 0.245 & 0.121 & --- \\
        \hline\hline
    \end{tabular}
    \caption{Marginal statistics of the five-mode model (Fig.~\ref{fig:five-marg}) before and after the affine map of Eq.~\eqref{eq:affine}. TVD is the total variation distance between histograms of target and generated samples, and $D$ and $p$ are the two-sample Kolmogorov--Smirnov statistic and $p$-value. The last row averages TVD and $D$ over the five assets.}
    \label{tab:marginals}
\end{table}

\emph{Twenty-mode residuals.} Table~\ref{tab:residuals} gives the distribution of residual magnitudes $|\Delta C_{ij}|$ over the 190 off-diagonal pairs of the twenty-mode model of Fig.~\ref{fig:twenty-corr}, averaged over five random initializations. More than half of the pairs have residuals below $0.05$, $86\%$ lie below $0.10$, and a single pair on average lies between $0.20$ and $0.50$. The seed-to-seed spread of each fraction is about two percentage points.

\begin{table}[t]
    \centering
    \footnotesize
    \begin{tabular}{@{}lccc@{}}
        \hline\hline
        Residual $|\Delta C_{ij}|$ & Pairs (mean) & Fraction & Cumulative \\
        \hline
        $[0,\,0.05)$    & 111 & $58.5\% \pm 2.1\%$ & $58.5\%$ \\
        $[0.05,\,0.10)$ & 53  & $27.7\% \pm 1.6\%$ & $86.2\%$ \\
        $[0.10,\,0.20)$ & 25  & $13.1\% \pm 2.1\%$ & $99.3\%$ \\
        $[0.20,\,0.50)$ & 1   & $0.7\% \pm 0.7\%$  & $100\%$ \\
        \hline\hline
    \end{tabular}
    \caption{Residual magnitudes $|\Delta C_{ij}|$ over the 190 off-diagonal pairs of the twenty-mode model (Fig.~\ref{fig:twenty-corr}). Fractions are means over five random initializations with the standard deviation across seeds; pair counts are the mean fractions multiplied by 190 and rounded.}
    \label{tab:residuals}
\end{table}

\begin{figure}
    \centering
    \includegraphics[width=0.92\columnwidth]{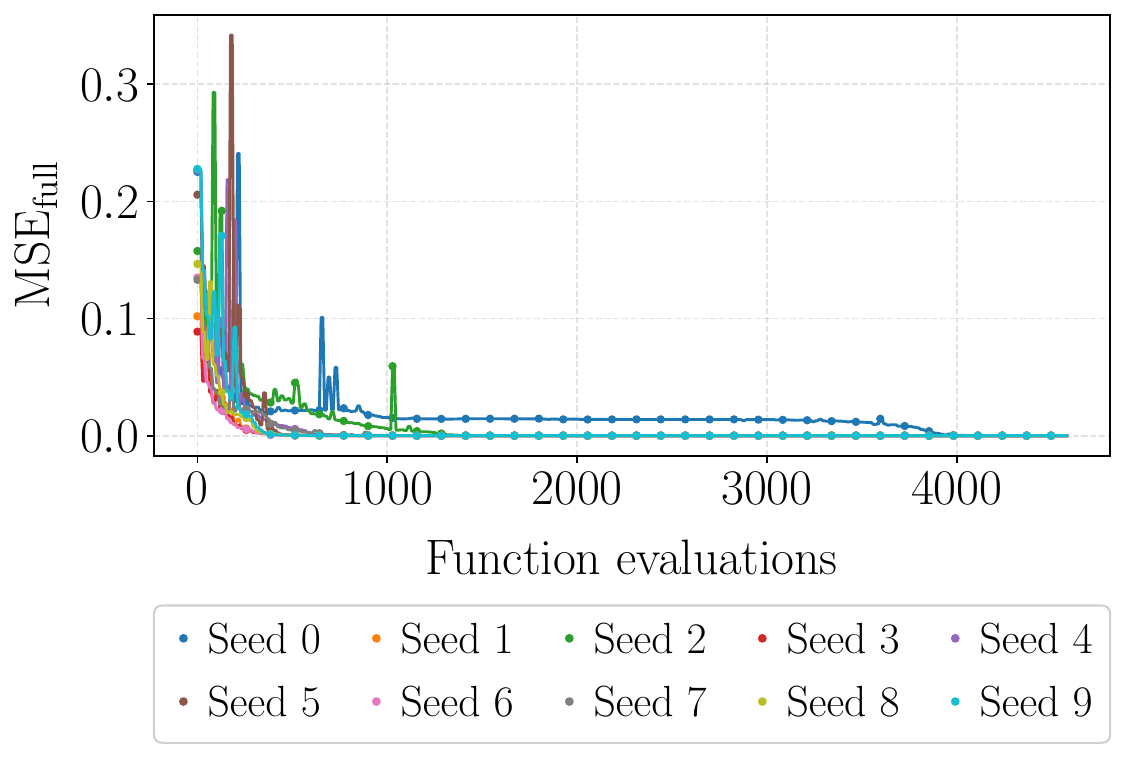}
    \caption{Convergence of the five-mode training. $\mathrm{MSE}_{\mathrm{full}}$ between $\Cgen(\bm{\theta})$ and $\Ctgt$ against the number of function evaluations, one curve per random initialization (ten seeds, see legend). Nine runs fall below $10^{-2}$ within about 860 evaluations and reach a common plateau close to zero, and all ten end below $3.1\times10^{-8}$. Isolated spikes are transient excursions during line searches.}
    \label{fig:five-conv}
\end{figure}

\begin{figure}
    \centering
    \includegraphics[width=0.92\columnwidth]{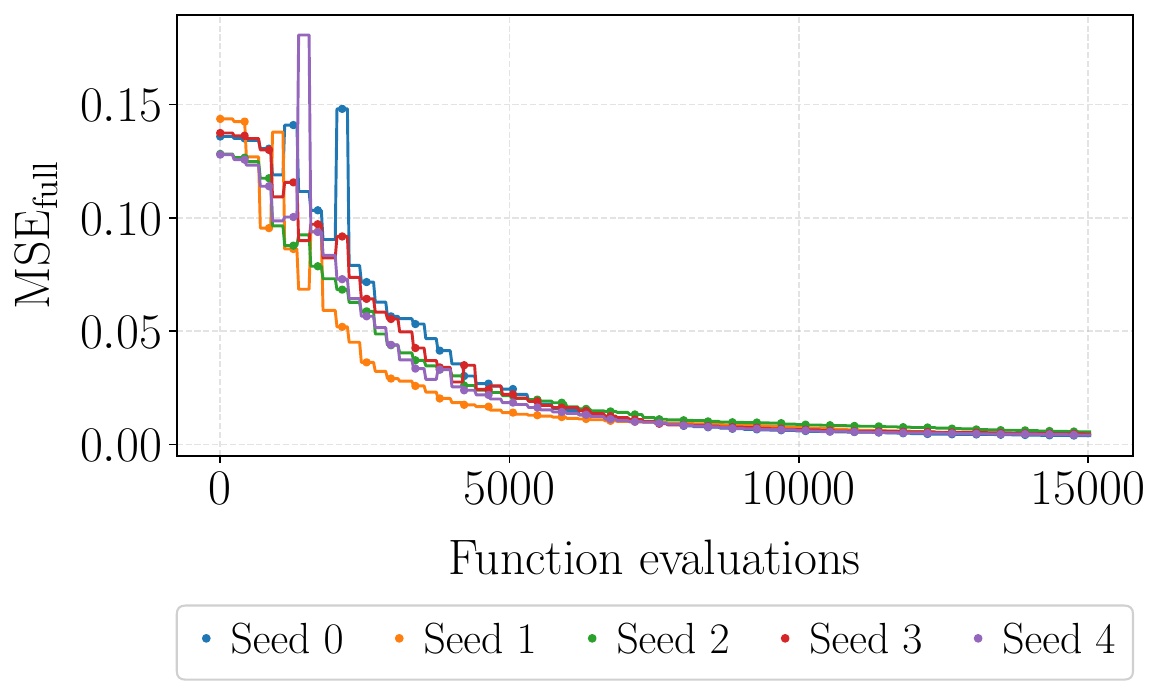}
    \caption{Convergence of the twenty-mode training. $\mathrm{MSE}_{\mathrm{full}}$ between $\Cgen(\bm{\theta})$ and $\Ctgt$ against the number of function evaluations, one curve per random initialization (five seeds, see legend). The traces start between $0.12$ and $0.15$, cross $0.05$ after roughly 4000 evaluations, and approach a common plateau below $0.01$ after roughly 10000 evaluations.}
    \label{fig:twenty-conv}
\end{figure}

\emph{Convergence.} Figs.~\ref{fig:five-conv} and \ref{fig:twenty-conv} show $\mathrm{MSE}_{\mathrm{full}}$ between $\Cgen(\bm{\theta})$ and $\Ctgt$ during training. Nine of the ten five-mode runs fall below $10^{-2}$ within $860$ function evaluations and below $10^{-6}$ within $1960$, with transient excursions of individual seeds during line searches. The tenth run remains near $1.4\times10^{-2}$ for about 3500 evaluations before it follows the others, and falls below $10^{-2}$ after 3554 and below $10^{-6}$ after 4276 evaluations. All ten runs end between $5.0\times10^{-10}$ and $3.0\times10^{-8}$, and the median is $4.3\times10^{-9}$. The twenty-mode traces decrease from initial values of $0.12$ to $0.15$, cross $0.05$ after roughly 4000 evaluations, and approach a common plateau below $0.01$ after roughly 10000 evaluations. At both sizes the runs end at comparable values, which supports the multi-seed statements in Secs.~\ref{sec:five-mode} and \ref{sec:twenty-mode}.

\FloatBarrier
\bibliographystyle{apsrev4-2}
\bibliography{references_v1.0}

\begin{thebibliography}{100}%
\makeatletter
\providecommand \@ifxundefined [1]{%
 \@ifx{#1\undefined}
}%
\providecommand \@ifnum [1]{%
 \ifnum #1\expandafter \@firstoftwo
 \else \expandafter \@secondoftwo
 \fi
}%
\providecommand \@ifx [1]{%
 \ifx #1\expandafter \@firstoftwo
 \else \expandafter \@secondoftwo
 \fi
}%
\providecommand \natexlab [1]{#1}%
\providecommand \enquote  [1]{``#1''}%
\providecommand \bibnamefont  [1]{#1}%
\providecommand \bibfnamefont [1]{#1}%
\providecommand \citenamefont [1]{#1}%
\providecommand \href@noop [0]{\@secondoftwo}%
\providecommand \href [0]{\begingroup \@sanitize@url \@href}%
\providecommand \@href[1]{\@@startlink{#1}\@@href}%
\providecommand \@@href[1]{\endgroup#1\@@endlink}%
\providecommand \@sanitize@url [0]{\catcode `\\12\catcode `\$12\catcode `\&12\catcode `\#12\catcode `\^12\catcode `\_12\catcode `\%12\relax}%
\providecommand \@@startlink[1]{}%
\providecommand \@@endlink[0]{}%
\providecommand \url  [0]{\begingroup\@sanitize@url \@url }%
\providecommand \@url [1]{\endgroup\@href {#1}{\urlprefix }}%
\providecommand \urlprefix  [0]{URL }%
\providecommand \Eprint [0]{\href }%
\providecommand \doibase [0]{https://doi.org/}%
\providecommand \selectlanguage [0]{\@gobble}%
\providecommand \bibinfo  [0]{\@secondoftwo}%
\providecommand \bibfield  [0]{\@secondoftwo}%
\providecommand \translation [1]{[#1]}%
\providecommand \BibitemOpen [0]{}%
\providecommand \bibitemStop [0]{}%
\providecommand \bibitemNoStop [0]{.\EOS\space}%
\providecommand \EOS [0]{\spacefactor3000\relax}%
\providecommand \BibitemShut  [1]{\csname bibitem#1\endcsname}%
\let\auto@bib@innerbib\@empty
\bibitem [{\citenamefont {McNeil}\ \emph {et~al.}(2005)\citenamefont {McNeil}, \citenamefont {Frey},\ and\ \citenamefont {Embrechts}}]{McNeil2005QRM}%
  \BibitemOpen
  \bibfield  {author} {\bibinfo {author} {\bibfnamefont {A.~J.}\ \bibnamefont {McNeil}}, \bibinfo {author} {\bibfnamefont {R.}~\bibnamefont {Frey}},\ and\ \bibinfo {author} {\bibfnamefont {P.}~\bibnamefont {Embrechts}},\ }\href@noop {} {\emph {\bibinfo {title} {Quantitative Risk Management: Concepts, Techniques and Tools}}}\ (\bibinfo  {publisher} {Princeton University Press},\ \bibinfo {address} {Princeton, NJ},\ \bibinfo {year} {2005})\BibitemShut {NoStop}%
\bibitem [{\citenamefont {Goodfellow}\ \emph {et~al.}(2016)\citenamefont {Goodfellow}, \citenamefont {Bengio},\ and\ \citenamefont {Courville}}]{Goodfellow2016DL}%
  \BibitemOpen
  \bibfield  {author} {\bibinfo {author} {\bibfnamefont {I.}~\bibnamefont {Goodfellow}}, \bibinfo {author} {\bibfnamefont {Y.}~\bibnamefont {Bengio}},\ and\ \bibinfo {author} {\bibfnamefont {A.}~\bibnamefont {Courville}},\ }\href@noop {} {\emph {\bibinfo {title} {Deep Learning}}}\ (\bibinfo  {publisher} {MIT Press},\ \bibinfo {address} {Cambridge, MA},\ \bibinfo {year} {2016})\BibitemShut {NoStop}%
\bibitem [{\citenamefont {Kingma}\ and\ \citenamefont {Welling}(2014)}]{Kingma2014VAE}%
  \BibitemOpen
  \bibfield  {author} {\bibinfo {author} {\bibfnamefont {D.~P.}\ \bibnamefont {Kingma}}\ and\ \bibinfo {author} {\bibfnamefont {M.}~\bibnamefont {Welling}},\ }in\ \href@noop {} {\emph {\bibinfo {booktitle} {International Conference on Learning Representations}}}\ (\bibinfo {year} {2014})\BibitemShut {NoStop}%
\bibitem [{\citenamefont {Goodfellow}\ \emph {et~al.}(2014)\citenamefont {Goodfellow}, \citenamefont {Pouget-Abadie}, \citenamefont {Mirza}, \citenamefont {Xu}, \citenamefont {Warde-Farley}, \citenamefont {Ozair}, \citenamefont {Courville},\ and\ \citenamefont {Bengio}}]{Goodfellow2014GAN}%
  \BibitemOpen
  \bibfield  {author} {\bibinfo {author} {\bibfnamefont {I.}~\bibnamefont {Goodfellow}}, \bibinfo {author} {\bibfnamefont {J.}~\bibnamefont {Pouget-Abadie}}, \bibinfo {author} {\bibfnamefont {M.}~\bibnamefont {Mirza}}, \bibinfo {author} {\bibfnamefont {B.}~\bibnamefont {Xu}}, \bibinfo {author} {\bibfnamefont {D.}~\bibnamefont {Warde-Farley}}, \bibinfo {author} {\bibfnamefont {S.}~\bibnamefont {Ozair}}, \bibinfo {author} {\bibfnamefont {A.}~\bibnamefont {Courville}},\ and\ \bibinfo {author} {\bibfnamefont {Y.}~\bibnamefont {Bengio}},\ }in\ \href@noop {} {\emph {\bibinfo {booktitle} {Advances in Neural Information Processing Systems}}},\ Vol.~\bibinfo {volume} {27}\ (\bibinfo {year} {2014})\BibitemShut {NoStop}%
\bibitem [{\citenamefont {Dinh}\ \emph {et~al.}(2017)\citenamefont {Dinh}, \citenamefont {Sohl-Dickstein},\ and\ \citenamefont {Bengio}}]{Dinh2017RealNVP}%
  \BibitemOpen
  \bibfield  {author} {\bibinfo {author} {\bibfnamefont {L.}~\bibnamefont {Dinh}}, \bibinfo {author} {\bibfnamefont {J.}~\bibnamefont {Sohl-Dickstein}},\ and\ \bibinfo {author} {\bibfnamefont {S.}~\bibnamefont {Bengio}},\ }in\ \href@noop {} {\emph {\bibinfo {booktitle} {International Conference on Learning Representations}}}\ (\bibinfo {year} {2017})\BibitemShut {NoStop}%
\bibitem [{\citenamefont {Papamakarios}\ \emph {et~al.}(2021)\citenamefont {Papamakarios}, \citenamefont {Nalisnick}, \citenamefont {Rezende}, \citenamefont {Mohamed},\ and\ \citenamefont {Lakshminarayanan}}]{Papamakarios2021NormalizingFlows}%
  \BibitemOpen
  \bibfield  {author} {\bibinfo {author} {\bibfnamefont {G.}~\bibnamefont {Papamakarios}}, \bibinfo {author} {\bibfnamefont {E.}~\bibnamefont {Nalisnick}}, \bibinfo {author} {\bibfnamefont {D.~J.}\ \bibnamefont {Rezende}}, \bibinfo {author} {\bibfnamefont {S.}~\bibnamefont {Mohamed}},\ and\ \bibinfo {author} {\bibfnamefont {B.}~\bibnamefont {Lakshminarayanan}},\ }\href@noop {} {\bibfield  {journal} {\bibinfo  {journal} {Journal of Machine Learning Research}\ }\textbf {\bibinfo {volume} {22}},\ \bibinfo {pages} {1} (\bibinfo {year} {2021})}\BibitemShut {NoStop}%
\bibitem [{\citenamefont {Ho}\ \emph {et~al.}(2020)\citenamefont {Ho}, \citenamefont {Jain},\ and\ \citenamefont {Abbeel}}]{Ho2020DDPM}%
  \BibitemOpen
  \bibfield  {author} {\bibinfo {author} {\bibfnamefont {J.}~\bibnamefont {Ho}}, \bibinfo {author} {\bibfnamefont {A.}~\bibnamefont {Jain}},\ and\ \bibinfo {author} {\bibfnamefont {P.}~\bibnamefont {Abbeel}},\ }in\ \href@noop {} {\emph {\bibinfo {booktitle} {Advances in Neural Information Processing Systems}}},\ Vol.~\bibinfo {volume} {33}\ (\bibinfo {year} {2020})\ pp.\ \bibinfo {pages} {6840--6851},\ \Eprint {https://arxiv.org/abs/2006.11239} {arXiv:2006.11239 [cs.LG]} \BibitemShut {NoStop}%
\bibitem [{\citenamefont {Song}\ \emph {et~al.}(2021)\citenamefont {Song}, \citenamefont {Sohl-Dickstein}, \citenamefont {Kingma}, \citenamefont {Kumar}, \citenamefont {Ermon},\ and\ \citenamefont {Poole}}]{Song2021ScoreSDE}%
  \BibitemOpen
  \bibfield  {author} {\bibinfo {author} {\bibfnamefont {Y.}~\bibnamefont {Song}}, \bibinfo {author} {\bibfnamefont {J.}~\bibnamefont {Sohl-Dickstein}}, \bibinfo {author} {\bibfnamefont {D.~P.}\ \bibnamefont {Kingma}}, \bibinfo {author} {\bibfnamefont {A.}~\bibnamefont {Kumar}}, \bibinfo {author} {\bibfnamefont {S.}~\bibnamefont {Ermon}},\ and\ \bibinfo {author} {\bibfnamefont {B.}~\bibnamefont {Poole}},\ }in\ \href@noop {} {\emph {\bibinfo {booktitle} {International Conference on Learning Representations}}}\ (\bibinfo {year} {2021})\ \Eprint {https://arxiv.org/abs/2011.13456} {arXiv:2011.13456 [cs.LG]} \BibitemShut {NoStop}%
\bibitem [{\citenamefont {McMahon}\ \emph {et~al.}(2016)\citenamefont {McMahon}, \citenamefont {Marandi}, \citenamefont {Haribara}, \citenamefont {Hamerly}, \citenamefont {Langrock}, \citenamefont {Tamate}, \citenamefont {Inagaki}, \citenamefont {Takesue}, \citenamefont {Utsunomiya}, \citenamefont {Aihara}, \citenamefont {Byer}, \citenamefont {Fejer}, \citenamefont {Mabuchi},\ and\ \citenamefont {Yamamoto}}]{McMahon2016CIM}%
  \BibitemOpen
  \bibfield  {author} {\bibinfo {author} {\bibfnamefont {P.~L.}\ \bibnamefont {McMahon}}, \bibinfo {author} {\bibfnamefont {A.}~\bibnamefont {Marandi}}, \bibinfo {author} {\bibfnamefont {Y.}~\bibnamefont {Haribara}}, \bibinfo {author} {\bibfnamefont {R.}~\bibnamefont {Hamerly}}, \bibinfo {author} {\bibfnamefont {C.}~\bibnamefont {Langrock}}, \bibinfo {author} {\bibfnamefont {S.}~\bibnamefont {Tamate}}, \bibinfo {author} {\bibfnamefont {T.}~\bibnamefont {Inagaki}}, \bibinfo {author} {\bibfnamefont {H.}~\bibnamefont {Takesue}}, \bibinfo {author} {\bibfnamefont {S.}~\bibnamefont {Utsunomiya}}, \bibinfo {author} {\bibfnamefont {K.}~\bibnamefont {Aihara}}, \bibinfo {author} {\bibfnamefont {R.~L.}\ \bibnamefont {Byer}}, \bibinfo {author} {\bibfnamefont {M.~M.}\ \bibnamefont {Fejer}}, \bibinfo {author} {\bibfnamefont {H.}~\bibnamefont {Mabuchi}},\ and\ \bibinfo {author} {\bibfnamefont {Y.}~\bibnamefont {Yamamoto}},\ }\href {https://doi.org/10.1126/science.aah5178} {\bibfield  {journal} {\bibinfo  {journal}
  {Science}\ }\textbf {\bibinfo {volume} {354}},\ \bibinfo {pages} {614} (\bibinfo {year} {2016})}\BibitemShut {NoStop}%
\bibitem [{\citenamefont {Inagaki}\ \emph {et~al.}(2016)\citenamefont {Inagaki}, \citenamefont {Haribara}, \citenamefont {Igarashi}, \citenamefont {Sonobe}, \citenamefont {Tamate}, \citenamefont {Honjo}, \citenamefont {Marandi}, \citenamefont {McMahon}, \citenamefont {Umeki}, \citenamefont {Enbutsu}, \citenamefont {Tadanaga}, \citenamefont {Takenouchi}, \citenamefont {Aihara}, \citenamefont {Kawarabayashi}, \citenamefont {Inoue}, \citenamefont {Utsunomiya},\ and\ \citenamefont {Takesue}}]{Inagaki2016CIM}%
  \BibitemOpen
  \bibfield  {author} {\bibinfo {author} {\bibfnamefont {T.}~\bibnamefont {Inagaki}}, \bibinfo {author} {\bibfnamefont {Y.}~\bibnamefont {Haribara}}, \bibinfo {author} {\bibfnamefont {K.}~\bibnamefont {Igarashi}}, \bibinfo {author} {\bibfnamefont {T.}~\bibnamefont {Sonobe}}, \bibinfo {author} {\bibfnamefont {S.}~\bibnamefont {Tamate}}, \bibinfo {author} {\bibfnamefont {T.}~\bibnamefont {Honjo}}, \bibinfo {author} {\bibfnamefont {A.}~\bibnamefont {Marandi}}, \bibinfo {author} {\bibfnamefont {P.~L.}\ \bibnamefont {McMahon}}, \bibinfo {author} {\bibfnamefont {T.}~\bibnamefont {Umeki}}, \bibinfo {author} {\bibfnamefont {K.}~\bibnamefont {Enbutsu}}, \bibinfo {author} {\bibfnamefont {O.}~\bibnamefont {Tadanaga}}, \bibinfo {author} {\bibfnamefont {H.}~\bibnamefont {Takenouchi}}, \bibinfo {author} {\bibfnamefont {K.}~\bibnamefont {Aihara}}, \bibinfo {author} {\bibfnamefont {K.-i.}\ \bibnamefont {Kawarabayashi}}, \bibinfo {author} {\bibfnamefont {K.}~\bibnamefont {Inoue}}, \bibinfo {author} {\bibfnamefont
  {S.}~\bibnamefont {Utsunomiya}},\ and\ \bibinfo {author} {\bibfnamefont {H.}~\bibnamefont {Takesue}},\ }\href {https://doi.org/10.1126/science.aah4243} {\bibfield  {journal} {\bibinfo  {journal} {Science}\ }\textbf {\bibinfo {volume} {354}},\ \bibinfo {pages} {603} (\bibinfo {year} {2016})}\BibitemShut {NoStop}%
\bibitem [{\citenamefont {Borders}\ \emph {et~al.}(2019)\citenamefont {Borders}, \citenamefont {Pervaiz}, \citenamefont {Fukami}, \citenamefont {Camsari}, \citenamefont {Ohno},\ and\ \citenamefont {Datta}}]{Borders2019pbits}%
  \BibitemOpen
  \bibfield  {author} {\bibinfo {author} {\bibfnamefont {W.~A.}\ \bibnamefont {Borders}}, \bibinfo {author} {\bibfnamefont {A.~Z.}\ \bibnamefont {Pervaiz}}, \bibinfo {author} {\bibfnamefont {S.}~\bibnamefont {Fukami}}, \bibinfo {author} {\bibfnamefont {K.~Y.}\ \bibnamefont {Camsari}}, \bibinfo {author} {\bibfnamefont {H.}~\bibnamefont {Ohno}},\ and\ \bibinfo {author} {\bibfnamefont {S.}~\bibnamefont {Datta}},\ }\href {https://doi.org/10.1038/s41586-019-1557-9} {\bibfield  {journal} {\bibinfo  {journal} {Nature}\ }\textbf {\bibinfo {volume} {573}},\ \bibinfo {pages} {390} (\bibinfo {year} {2019})}\BibitemShut {NoStop}%
\bibitem [{\citenamefont {Coles}\ \emph {et~al.}(2023)\citenamefont {Coles}, \citenamefont {Szczepanski}, \citenamefont {Melanson}, \citenamefont {Donatella}, \citenamefont {Martinez},\ and\ \citenamefont {Sbahi}}]{Coles2023ThermodynamicAI}%
  \BibitemOpen
  \bibfield  {author} {\bibinfo {author} {\bibfnamefont {P.~J.}\ \bibnamefont {Coles}}, \bibinfo {author} {\bibfnamefont {C.}~\bibnamefont {Szczepanski}}, \bibinfo {author} {\bibfnamefont {D.}~\bibnamefont {Melanson}}, \bibinfo {author} {\bibfnamefont {K.}~\bibnamefont {Donatella}}, \bibinfo {author} {\bibfnamefont {A.~J.}\ \bibnamefont {Martinez}},\ and\ \bibinfo {author} {\bibfnamefont {F.}~\bibnamefont {Sbahi}},\ }in\ \href {https://doi.org/10.1109/ICRC60800.2023.10386858} {\emph {\bibinfo {booktitle} {2023 IEEE International Conference on Rebooting Computing (ICRC)}}}\ (\bibinfo {year} {2023})\ pp.\ \bibinfo {pages} {1--10},\ \Eprint {https://arxiv.org/abs/2302.06584} {arXiv:2302.06584 [cs.ET]} \BibitemShut {NoStop}%
\bibitem [{\citenamefont {Shen}\ \emph {et~al.}(2017)\citenamefont {Shen}, \citenamefont {Harris}, \citenamefont {Skirlo}, \citenamefont {Prabhu}, \citenamefont {Baehr-Jones}, \citenamefont {Hochberg}, \citenamefont {Sun}, \citenamefont {Zhao}, \citenamefont {Larochelle}, \citenamefont {Englund},\ and\ \citenamefont {Solja{\v{c}}i{\'c}}}]{Shen2017NanophotonicDL}%
  \BibitemOpen
  \bibfield  {author} {\bibinfo {author} {\bibfnamefont {Y.}~\bibnamefont {Shen}}, \bibinfo {author} {\bibfnamefont {N.~C.}\ \bibnamefont {Harris}}, \bibinfo {author} {\bibfnamefont {S.}~\bibnamefont {Skirlo}}, \bibinfo {author} {\bibfnamefont {M.}~\bibnamefont {Prabhu}}, \bibinfo {author} {\bibfnamefont {T.}~\bibnamefont {Baehr-Jones}}, \bibinfo {author} {\bibfnamefont {M.}~\bibnamefont {Hochberg}}, \bibinfo {author} {\bibfnamefont {X.}~\bibnamefont {Sun}}, \bibinfo {author} {\bibfnamefont {S.}~\bibnamefont {Zhao}}, \bibinfo {author} {\bibfnamefont {H.}~\bibnamefont {Larochelle}}, \bibinfo {author} {\bibfnamefont {D.}~\bibnamefont {Englund}},\ and\ \bibinfo {author} {\bibfnamefont {M.}~\bibnamefont {Solja{\v{c}}i{\'c}}},\ }\href {https://doi.org/10.1038/nphoton.2017.93} {\bibfield  {journal} {\bibinfo  {journal} {Nature Photonics}\ }\textbf {\bibinfo {volume} {11}},\ \bibinfo {pages} {441} (\bibinfo {year} {2017})}\BibitemShut {NoStop}%
\bibitem [{\citenamefont {Wetzstein}\ \emph {et~al.}(2020)\citenamefont {Wetzstein}, \citenamefont {Ozcan}, \citenamefont {Gigan}, \citenamefont {Fan}, \citenamefont {Englund}, \citenamefont {Solja{\v{c}}i{\'c}}, \citenamefont {Denz}, \citenamefont {Miller},\ and\ \citenamefont {Psaltis}}]{Wetzstein2020InferencePhotonics}%
  \BibitemOpen
  \bibfield  {author} {\bibinfo {author} {\bibfnamefont {G.}~\bibnamefont {Wetzstein}}, \bibinfo {author} {\bibfnamefont {A.}~\bibnamefont {Ozcan}}, \bibinfo {author} {\bibfnamefont {S.}~\bibnamefont {Gigan}}, \bibinfo {author} {\bibfnamefont {S.}~\bibnamefont {Fan}}, \bibinfo {author} {\bibfnamefont {D.}~\bibnamefont {Englund}}, \bibinfo {author} {\bibfnamefont {M.}~\bibnamefont {Solja{\v{c}}i{\'c}}}, \bibinfo {author} {\bibfnamefont {C.}~\bibnamefont {Denz}}, \bibinfo {author} {\bibfnamefont {D.~A.~B.}\ \bibnamefont {Miller}},\ and\ \bibinfo {author} {\bibfnamefont {D.}~\bibnamefont {Psaltis}},\ }\href {https://doi.org/10.1038/s41586-020-2973-6} {\bibfield  {journal} {\bibinfo  {journal} {Nature}\ }\textbf {\bibinfo {volume} {588}},\ \bibinfo {pages} {39} (\bibinfo {year} {2020})}\BibitemShut {NoStop}%
\bibitem [{\citenamefont {Wang}\ \emph {et~al.}(2026)\citenamefont {Wang}, \citenamefont {Muszynski}, \citenamefont {Dash}, \citenamefont {Kaurav}, \citenamefont {Menon},\ and\ \citenamefont {Kyriienko}}]{Wang2026PolaritonGAN}%
  \BibitemOpen
  \bibfield  {author} {\bibinfo {author} {\bibfnamefont {Y.}~\bibnamefont {Wang}}, \bibinfo {author} {\bibfnamefont {M.}~\bibnamefont {Muszynski}}, \bibinfo {author} {\bibfnamefont {A.}~\bibnamefont {Dash}}, \bibinfo {author} {\bibfnamefont {R.}~\bibnamefont {Kaurav}}, \bibinfo {author} {\bibfnamefont {V.~M.}\ \bibnamefont {Menon}},\ and\ \bibinfo {author} {\bibfnamefont {O.}~\bibnamefont {Kyriienko}},\ }\href@noop {} {\bibinfo {title} {Generative modelling powered by room-temperature polariton condensates}} (\bibinfo {year} {2026}),\ \Eprint {https://arxiv.org/abs/2606.15344} {arXiv:2606.15344 [cond-mat.dis-nn]} \BibitemShut {NoStop}%
\bibitem [{\citenamefont {Liu}\ and\ \citenamefont {Wang}(2018)}]{Liu2018QCBM}%
  \BibitemOpen
  \bibfield  {author} {\bibinfo {author} {\bibfnamefont {J.-G.}\ \bibnamefont {Liu}}\ and\ \bibinfo {author} {\bibfnamefont {L.}~\bibnamefont {Wang}},\ }\href {https://doi.org/10.1103/PhysRevA.98.062324} {\bibfield  {journal} {\bibinfo  {journal} {Physical Review A}\ }\textbf {\bibinfo {volume} {98}},\ \bibinfo {pages} {062324} (\bibinfo {year} {2018})}\BibitemShut {NoStop}%
\bibitem [{\citenamefont {Benedetti}\ \emph {et~al.}(2019)\citenamefont {Benedetti}, \citenamefont {Lloyd}, \citenamefont {Sack},\ and\ \citenamefont {Fiorentini}}]{Benedetti2019QML}%
  \BibitemOpen
  \bibfield  {author} {\bibinfo {author} {\bibfnamefont {M.}~\bibnamefont {Benedetti}}, \bibinfo {author} {\bibfnamefont {E.}~\bibnamefont {Lloyd}}, \bibinfo {author} {\bibfnamefont {S.}~\bibnamefont {Sack}},\ and\ \bibinfo {author} {\bibfnamefont {M.}~\bibnamefont {Fiorentini}},\ }\href {https://doi.org/10.1088/2058-9565/ab4eb5} {\bibfield  {journal} {\bibinfo  {journal} {Quantum Science and Technology}\ }\textbf {\bibinfo {volume} {4}},\ \bibinfo {pages} {043001} (\bibinfo {year} {2019})}\BibitemShut {NoStop}%
\bibitem [{\citenamefont {Tian}\ \emph {et~al.}(2023)\citenamefont {Tian}, \citenamefont {Sun}, \citenamefont {Du}, \citenamefont {Zhao}, \citenamefont {Liu}, \citenamefont {Zhang}, \citenamefont {Yi}, \citenamefont {Huang}, \citenamefont {Wang}, \citenamefont {Wu}, \citenamefont {Hsieh}, \citenamefont {Liu}, \citenamefont {Yang},\ and\ \citenamefont {Tao}}]{Tian2023QGMReview}%
  \BibitemOpen
  \bibfield  {author} {\bibinfo {author} {\bibfnamefont {J.}~\bibnamefont {Tian}}, \bibinfo {author} {\bibfnamefont {X.}~\bibnamefont {Sun}}, \bibinfo {author} {\bibfnamefont {Y.}~\bibnamefont {Du}}, \bibinfo {author} {\bibfnamefont {S.}~\bibnamefont {Zhao}}, \bibinfo {author} {\bibfnamefont {Q.}~\bibnamefont {Liu}}, \bibinfo {author} {\bibfnamefont {K.}~\bibnamefont {Zhang}}, \bibinfo {author} {\bibfnamefont {W.}~\bibnamefont {Yi}}, \bibinfo {author} {\bibfnamefont {W.}~\bibnamefont {Huang}}, \bibinfo {author} {\bibfnamefont {C.}~\bibnamefont {Wang}}, \bibinfo {author} {\bibfnamefont {X.}~\bibnamefont {Wu}}, \bibinfo {author} {\bibfnamefont {M.-H.}\ \bibnamefont {Hsieh}}, \bibinfo {author} {\bibfnamefont {T.}~\bibnamefont {Liu}}, \bibinfo {author} {\bibfnamefont {W.}~\bibnamefont {Yang}},\ and\ \bibinfo {author} {\bibfnamefont {D.}~\bibnamefont {Tao}},\ }\href {https://doi.org/10.1109/TPAMI.2023.3272029} {\bibfield  {journal} {\bibinfo  {journal} {IEEE Transactions on Pattern Analysis and Machine
  Intelligence}\ }\textbf {\bibinfo {volume} {45}},\ \bibinfo {pages} {12321} (\bibinfo {year} {2023})}\BibitemShut {NoStop}%
\bibitem [{\citenamefont {Lloyd}\ and\ \citenamefont {Weedbrook}(2018)}]{Lloyd2018QGAN}%
  \BibitemOpen
  \bibfield  {author} {\bibinfo {author} {\bibfnamefont {S.}~\bibnamefont {Lloyd}}\ and\ \bibinfo {author} {\bibfnamefont {C.}~\bibnamefont {Weedbrook}},\ }\href {https://doi.org/10.1103/PhysRevLett.121.040502} {\bibfield  {journal} {\bibinfo  {journal} {Physical Review Letters}\ }\textbf {\bibinfo {volume} {121}},\ \bibinfo {pages} {040502} (\bibinfo {year} {2018})}\BibitemShut {NoStop}%
\bibitem [{\citenamefont {Zoufal}\ \emph {et~al.}(2019)\citenamefont {Zoufal}, \citenamefont {Lucchi},\ and\ \citenamefont {Woerner}}]{Zoufal2019QGAN}%
  \BibitemOpen
  \bibfield  {author} {\bibinfo {author} {\bibfnamefont {C.}~\bibnamefont {Zoufal}}, \bibinfo {author} {\bibfnamefont {A.}~\bibnamefont {Lucchi}},\ and\ \bibinfo {author} {\bibfnamefont {S.}~\bibnamefont {Woerner}},\ }\href {https://doi.org/10.1038/s41534-019-0223-2} {\bibfield  {journal} {\bibinfo  {journal} {npj Quantum Information}\ }\textbf {\bibinfo {volume} {5}},\ \bibinfo {pages} {103} (\bibinfo {year} {2019})}\BibitemShut {NoStop}%
\bibitem [{\citenamefont {Coyle}\ \emph {et~al.}(2020)\citenamefont {Coyle}, \citenamefont {Mills}, \citenamefont {Danos},\ and\ \citenamefont {Kashefi}}]{Coyle2020BornSupremacy}%
  \BibitemOpen
  \bibfield  {author} {\bibinfo {author} {\bibfnamefont {B.}~\bibnamefont {Coyle}}, \bibinfo {author} {\bibfnamefont {D.}~\bibnamefont {Mills}}, \bibinfo {author} {\bibfnamefont {V.}~\bibnamefont {Danos}},\ and\ \bibinfo {author} {\bibfnamefont {E.}~\bibnamefont {Kashefi}},\ }\href {https://doi.org/10.1038/s41534-020-00288-9} {\bibfield  {journal} {\bibinfo  {journal} {npj Quantum Information}\ }\textbf {\bibinfo {volume} {6}},\ \bibinfo {pages} {60} (\bibinfo {year} {2020})}\BibitemShut {NoStop}%
\bibitem [{\citenamefont {Du}\ \emph {et~al.}(2022)\citenamefont {Du}, \citenamefont {Tu}, \citenamefont {Wu}, \citenamefont {Yuan},\ and\ \citenamefont {Tao}}]{Du2022PowerQuantumGenerativeLearning}%
  \BibitemOpen
  \bibfield  {author} {\bibinfo {author} {\bibfnamefont {Y.}~\bibnamefont {Du}}, \bibinfo {author} {\bibfnamefont {Z.}~\bibnamefont {Tu}}, \bibinfo {author} {\bibfnamefont {B.}~\bibnamefont {Wu}}, \bibinfo {author} {\bibfnamefont {X.}~\bibnamefont {Yuan}},\ and\ \bibinfo {author} {\bibfnamefont {D.}~\bibnamefont {Tao}},\ }\href@noop {} {\bibinfo {title} {Power of quantum generative learning}} (\bibinfo {year} {2022}),\ \Eprint {https://arxiv.org/abs/2205.04730} {arXiv:2205.04730 [quant-ph]} \BibitemShut {NoStop}%
\bibitem [{\citenamefont {Kyriienko}\ \emph {et~al.}(2024)\citenamefont {Kyriienko}, \citenamefont {Paine},\ and\ \citenamefont {Elfving}}]{Kyriienko2024TrainableProtocols}%
  \BibitemOpen
  \bibfield  {author} {\bibinfo {author} {\bibfnamefont {O.}~\bibnamefont {Kyriienko}}, \bibinfo {author} {\bibfnamefont {A.~E.}\ \bibnamefont {Paine}},\ and\ \bibinfo {author} {\bibfnamefont {V.~E.}\ \bibnamefont {Elfving}},\ }\href {https://doi.org/10.1103/PhysRevResearch.6.033291} {\bibfield  {journal} {\bibinfo  {journal} {Physical Review Research}\ }\textbf {\bibinfo {volume} {6}},\ \bibinfo {pages} {033291} (\bibinfo {year} {2024})}\BibitemShut {NoStop}%
\bibitem [{\citenamefont {Rudolph}\ \emph {et~al.}(2024)\citenamefont {Rudolph}, \citenamefont {Lerch}, \citenamefont {Thanasilp}, \citenamefont {Kiss}, \citenamefont {Shaya}, \citenamefont {Vallecorsa}, \citenamefont {Grossi},\ and\ \citenamefont {Holmes}}]{Rudolph2024Trainability}%
  \BibitemOpen
  \bibfield  {author} {\bibinfo {author} {\bibfnamefont {M.~S.}\ \bibnamefont {Rudolph}}, \bibinfo {author} {\bibfnamefont {S.}~\bibnamefont {Lerch}}, \bibinfo {author} {\bibfnamefont {S.}~\bibnamefont {Thanasilp}}, \bibinfo {author} {\bibfnamefont {O.}~\bibnamefont {Kiss}}, \bibinfo {author} {\bibfnamefont {O.}~\bibnamefont {Shaya}}, \bibinfo {author} {\bibfnamefont {S.}~\bibnamefont {Vallecorsa}}, \bibinfo {author} {\bibfnamefont {M.}~\bibnamefont {Grossi}},\ and\ \bibinfo {author} {\bibfnamefont {Z.}~\bibnamefont {Holmes}},\ }\href {https://doi.org/10.1038/s41534-024-00902-0} {\bibfield  {journal} {\bibinfo  {journal} {npj Quantum Information}\ }\textbf {\bibinfo {volume} {10}},\ \bibinfo {pages} {116} (\bibinfo {year} {2024})}\BibitemShut {NoStop}%
\bibitem [{\citenamefont {Gili}\ \emph {et~al.}(2024)\citenamefont {Gili}, \citenamefont {Mauri},\ and\ \citenamefont {Perdomo-Ortiz}}]{Gili2023Generalization}%
  \BibitemOpen
  \bibfield  {author} {\bibinfo {author} {\bibfnamefont {K.}~\bibnamefont {Gili}}, \bibinfo {author} {\bibfnamefont {M.}~\bibnamefont {Mauri}},\ and\ \bibinfo {author} {\bibfnamefont {A.}~\bibnamefont {Perdomo-Ortiz}},\ }\href {https://doi.org/10.1103/PhysRevApplied.21.044032} {\bibfield  {journal} {\bibinfo  {journal} {Physical Review Applied}\ }\textbf {\bibinfo {volume} {21}},\ \bibinfo {pages} {044032} (\bibinfo {year} {2024})}\BibitemShut {NoStop}%
\bibitem [{\citenamefont {Aaronson}\ and\ \citenamefont {Arkhipov}(2011)}]{Aaronson2011BosonSampling}%
  \BibitemOpen
  \bibfield  {author} {\bibinfo {author} {\bibfnamefont {S.}~\bibnamefont {Aaronson}}\ and\ \bibinfo {author} {\bibfnamefont {A.}~\bibnamefont {Arkhipov}},\ }in\ \href {https://doi.org/10.1145/1993636.1993682} {\emph {\bibinfo {booktitle} {Proceedings of the forty-third annual ACM symposium on Theory of computing}}},\ \bibinfo {series and number} {STOC '11}\ (\bibinfo  {publisher} {ACM},\ \bibinfo {year} {2011})\ pp.\ \bibinfo {pages} {333--342}\BibitemShut {NoStop}%
\bibitem [{\citenamefont {Hamilton}\ \emph {et~al.}(2017)\citenamefont {Hamilton}, \citenamefont {Kruse}, \citenamefont {Sansoni}, \citenamefont {Barkhofen}, \citenamefont {Silberhorn},\ and\ \citenamefont {Jex}}]{Hamilton2017GBS}%
  \BibitemOpen
  \bibfield  {author} {\bibinfo {author} {\bibfnamefont {C.~S.}\ \bibnamefont {Hamilton}}, \bibinfo {author} {\bibfnamefont {R.}~\bibnamefont {Kruse}}, \bibinfo {author} {\bibfnamefont {L.}~\bibnamefont {Sansoni}}, \bibinfo {author} {\bibfnamefont {S.}~\bibnamefont {Barkhofen}}, \bibinfo {author} {\bibfnamefont {C.}~\bibnamefont {Silberhorn}},\ and\ \bibinfo {author} {\bibfnamefont {I.}~\bibnamefont {Jex}},\ }\href {https://doi.org/10.1103/PhysRevLett.119.170501} {\bibfield  {journal} {\bibinfo  {journal} {Physical Review Letters}\ }\textbf {\bibinfo {volume} {119}},\ \bibinfo {pages} {170501} (\bibinfo {year} {2017})}\BibitemShut {NoStop}%
\bibitem [{\citenamefont {Kolarovszki}\ \emph {et~al.}(2026)\citenamefont {Kolarovszki}, \citenamefont {Bak{\'o}}, \citenamefont {Oszmaniec}, \citenamefont {Oh},\ and\ \citenamefont {Zimbor{\'a}s}}]{Kolarovszki2026BosonicBornMachines}%
  \BibitemOpen
  \bibfield  {author} {\bibinfo {author} {\bibfnamefont {Z.}~\bibnamefont {Kolarovszki}}, \bibinfo {author} {\bibfnamefont {B.}~\bibnamefont {Bak{\'o}}}, \bibinfo {author} {\bibfnamefont {M.}~\bibnamefont {Oszmaniec}}, \bibinfo {author} {\bibfnamefont {C.}~\bibnamefont {Oh}},\ and\ \bibinfo {author} {\bibfnamefont {Z.}~\bibnamefont {Zimbor{\'a}s}},\ }\href@noop {} {\bibinfo {title} {Generative modeling with {G}aussian boson sampling: Classically trainable bosonic {B}orn machines}} (\bibinfo {year} {2026}),\ \Eprint {https://arxiv.org/abs/2603.11195} {arXiv:2603.11195 [quant-ph]} \BibitemShut {NoStop}%
\bibitem [{\citenamefont {Kurkin}\ \emph {et~al.}(2026)\citenamefont {Kurkin}, \citenamefont {Chabaud}, \citenamefont {Kolarovszki}, \citenamefont {Bak{\'o}}, \citenamefont {Zimbor{\'a}s},\ and\ \citenamefont {Dunjko}}]{Kurkin2026BosonSamplingGenerative}%
  \BibitemOpen
  \bibfield  {author} {\bibinfo {author} {\bibfnamefont {A.}~\bibnamefont {Kurkin}}, \bibinfo {author} {\bibfnamefont {U.}~\bibnamefont {Chabaud}}, \bibinfo {author} {\bibfnamefont {Z.}~\bibnamefont {Kolarovszki}}, \bibinfo {author} {\bibfnamefont {B.}~\bibnamefont {Bak{\'o}}}, \bibinfo {author} {\bibfnamefont {Z.}~\bibnamefont {Zimbor{\'a}s}},\ and\ \bibinfo {author} {\bibfnamefont {V.}~\bibnamefont {Dunjko}},\ }\href@noop {} {\bibinfo {title} {Universality of classically trainable, quantum-deployed boson-sampling generative models}} (\bibinfo {year} {2026}),\ \Eprint {https://arxiv.org/abs/2603.11014} {arXiv:2603.11014 [quant-ph]} \BibitemShut {NoStop}%
\bibitem [{\citenamefont {Oszmaniec}\ \emph {et~al.}(2022)\citenamefont {Oszmaniec}, \citenamefont {Dangniam}, \citenamefont {Morales},\ and\ \citenamefont {Zimbor{\'a}s}}]{Oszmaniec2022FermionSampling}%
  \BibitemOpen
  \bibfield  {author} {\bibinfo {author} {\bibfnamefont {M.}~\bibnamefont {Oszmaniec}}, \bibinfo {author} {\bibfnamefont {N.}~\bibnamefont {Dangniam}}, \bibinfo {author} {\bibfnamefont {M.~E.~S.}\ \bibnamefont {Morales}},\ and\ \bibinfo {author} {\bibfnamefont {Z.}~\bibnamefont {Zimbor{\'a}s}},\ }\href {https://doi.org/10.1103/PRXQuantum.3.020328} {\bibfield  {journal} {\bibinfo  {journal} {PRX Quantum}\ }\textbf {\bibinfo {volume} {3}},\ \bibinfo {pages} {020328} (\bibinfo {year} {2022})}\BibitemShut {NoStop}%
\bibitem [{\citenamefont {Bak{\'o}}\ \emph {et~al.}(2025)\citenamefont {Bak{\'o}}, \citenamefont {Kolarovszki},\ and\ \citenamefont {Zimbor{\'a}s}}]{bako2025fermionic}%
  \BibitemOpen
  \bibfield  {author} {\bibinfo {author} {\bibfnamefont {B.}~\bibnamefont {Bak{\'o}}}, \bibinfo {author} {\bibfnamefont {Z.}~\bibnamefont {Kolarovszki}},\ and\ \bibinfo {author} {\bibfnamefont {Z.}~\bibnamefont {Zimbor{\'a}s}},\ }\href@noop {} {\bibinfo {title} {Fermionic {B}orn machines: Classical training of quantum generative models based on {F}ermion {S}ampling}} (\bibinfo {year} {2025}),\ \Eprint {https://arxiv.org/abs/2511.13844} {arXiv:2511.13844 [quant-ph]} \BibitemShut {NoStop}%
\bibitem [{\citenamefont {Cheng}\ \emph {et~al.}(2018)\citenamefont {Cheng}, \citenamefont {Chen},\ and\ \citenamefont {Wang}}]{Cheng2018BornVsBoltzmann}%
  \BibitemOpen
  \bibfield  {author} {\bibinfo {author} {\bibfnamefont {S.}~\bibnamefont {Cheng}}, \bibinfo {author} {\bibfnamefont {J.}~\bibnamefont {Chen}},\ and\ \bibinfo {author} {\bibfnamefont {L.}~\bibnamefont {Wang}},\ }\href {https://doi.org/10.3390/e20080583} {\bibfield  {journal} {\bibinfo  {journal} {Entropy}\ }\textbf {\bibinfo {volume} {20}},\ \bibinfo {pages} {583} (\bibinfo {year} {2018})}\BibitemShut {NoStop}%
\bibitem [{\citenamefont {Amin}\ \emph {et~al.}(2018)\citenamefont {Amin}, \citenamefont {Andriyash}, \citenamefont {Rolfe}, \citenamefont {Kulchytskyy},\ and\ \citenamefont {Melko}}]{Amin2018QBM}%
  \BibitemOpen
  \bibfield  {author} {\bibinfo {author} {\bibfnamefont {M.~H.}\ \bibnamefont {Amin}}, \bibinfo {author} {\bibfnamefont {E.}~\bibnamefont {Andriyash}}, \bibinfo {author} {\bibfnamefont {J.}~\bibnamefont {Rolfe}}, \bibinfo {author} {\bibfnamefont {B.}~\bibnamefont {Kulchytskyy}},\ and\ \bibinfo {author} {\bibfnamefont {R.}~\bibnamefont {Melko}},\ }\href {https://doi.org/10.1103/PhysRevX.8.021050} {\bibfield  {journal} {\bibinfo  {journal} {Physical Review X}\ }\textbf {\bibinfo {volume} {8}},\ \bibinfo {pages} {021050} (\bibinfo {year} {2018})}\BibitemShut {NoStop}%
\bibitem [{\citenamefont {Kieferov{\'a}}\ and\ \citenamefont {Wiebe}(2017)}]{Kieferova2017QBM}%
  \BibitemOpen
  \bibfield  {author} {\bibinfo {author} {\bibfnamefont {M.}~\bibnamefont {Kieferov{\'a}}}\ and\ \bibinfo {author} {\bibfnamefont {N.}~\bibnamefont {Wiebe}},\ }\href {https://doi.org/10.1103/PhysRevA.96.062327} {\bibfield  {journal} {\bibinfo  {journal} {Physical Review A}\ }\textbf {\bibinfo {volume} {96}},\ \bibinfo {pages} {062327} (\bibinfo {year} {2017})}\BibitemShut {NoStop}%
\bibitem [{\citenamefont {Kasture}\ \emph {et~al.}(2023)\citenamefont {Kasture}, \citenamefont {Kyriienko},\ and\ \citenamefont {Elfving}}]{Kasture2023Protocols}%
  \BibitemOpen
  \bibfield  {author} {\bibinfo {author} {\bibfnamefont {S.}~\bibnamefont {Kasture}}, \bibinfo {author} {\bibfnamefont {O.}~\bibnamefont {Kyriienko}},\ and\ \bibinfo {author} {\bibfnamefont {V.~E.}\ \bibnamefont {Elfving}},\ }\href {https://doi.org/10.1103/PhysRevA.108.042406} {\bibfield  {journal} {\bibinfo  {journal} {Physical Review A}\ }\textbf {\bibinfo {volume} {108}},\ \bibinfo {pages} {042406} (\bibinfo {year} {2023})}\BibitemShut {NoStop}%
\bibitem [{\citenamefont {T{\"u}ys{\"u}z}\ \emph {et~al.}(2026)\citenamefont {T{\"u}ys{\"u}z}, \citenamefont {Kyriienko},\ and\ \citenamefont {Grossi}}]{Tuysuz2026Fourier}%
  \BibitemOpen
  \bibfield  {author} {\bibinfo {author} {\bibfnamefont {C.}~\bibnamefont {T{\"u}ys{\"u}z}}, \bibinfo {author} {\bibfnamefont {O.}~\bibnamefont {Kyriienko}},\ and\ \bibinfo {author} {\bibfnamefont {M.}~\bibnamefont {Grossi}},\ }\href@noop {} {\bibinfo {title} {Quantum {F}ourier generative models trainable at large scale}} (\bibinfo {year} {2026}),\ \Eprint {https://arxiv.org/abs/2606.28483} {arXiv:2606.28483 [quant-ph]} \BibitemShut {NoStop}%
\bibitem [{\citenamefont {Huang}\ \emph {et~al.}(2026)\citenamefont {Huang}, \citenamefont {Maxwell}, \citenamefont {Belis}, \citenamefont {Peters}, \citenamefont {Pye}, \citenamefont {Jahangiri},\ and\ \citenamefont {Bowles}}]{Huang2026SpectralBornMachines}%
  \BibitemOpen
  \bibfield  {author} {\bibinfo {author} {\bibfnamefont {A.}~\bibnamefont {Huang}}, \bibinfo {author} {\bibfnamefont {W.}~\bibnamefont {Maxwell}}, \bibinfo {author} {\bibfnamefont {V.}~\bibnamefont {Belis}}, \bibinfo {author} {\bibfnamefont {E.}~\bibnamefont {Peters}}, \bibinfo {author} {\bibfnamefont {J.}~\bibnamefont {Pye}}, \bibinfo {author} {\bibfnamefont {S.}~\bibnamefont {Jahangiri}},\ and\ \bibinfo {author} {\bibfnamefont {J.}~\bibnamefont {Bowles}},\ }\href@noop {} {\bibinfo {title} {Spectral {B}orn machines: Classically trainable quantum generative models for discrete data}} (\bibinfo {year} {2026}),\ \Eprint {https://arxiv.org/abs/2607.06675} {arXiv:2607.06675 [quant-ph]} \BibitemShut {NoStop}%
\bibitem [{\citenamefont {Mahassen}\ \emph {et~al.}(2026)\citenamefont {Mahassen}, \citenamefont {Jastrzebski},\ and\ \citenamefont {Kyriienko}}]{Mahassen2026FokkerPlanck}%
  \BibitemOpen
  \bibfield  {author} {\bibinfo {author} {\bibfnamefont {T.}~\bibnamefont {Mahassen}}, \bibinfo {author} {\bibfnamefont {M.}~\bibnamefont {Jastrzebski}},\ and\ \bibinfo {author} {\bibfnamefont {O.}~\bibnamefont {Kyriienko}},\ }\href@noop {} {\bibinfo {title} {Quantum {F}okker--{P}lanck solvers as trainable generative models}} (\bibinfo {year} {2026}),\ \Eprint {https://arxiv.org/abs/2610.31330} {arXiv:2610.31330 [quant-ph]} \BibitemShut {NoStop}%
\bibitem [{\citenamefont {Recio-Armengol}\ \emph {et~al.}(2025)\citenamefont {Recio-Armengol}, \citenamefont {Ahmed},\ and\ \citenamefont {Bowles}}]{RecioArmengol2025TrainClassicalDeployQuantum}%
  \BibitemOpen
  \bibfield  {author} {\bibinfo {author} {\bibfnamefont {E.}~\bibnamefont {Recio-Armengol}}, \bibinfo {author} {\bibfnamefont {S.}~\bibnamefont {Ahmed}},\ and\ \bibinfo {author} {\bibfnamefont {J.}~\bibnamefont {Bowles}},\ }\href@noop {} {\bibinfo {title} {Train on classical, deploy on quantum: scaling generative quantum machine learning to a thousand qubits}} (\bibinfo {year} {2025}),\ \Eprint {https://arxiv.org/abs/2503.02934} {arXiv:2503.02934 [quant-ph]} \BibitemShut {NoStop}%
\bibitem [{\citenamefont {Lerch}\ \emph {et~al.}(2026)\citenamefont {Lerch}, \citenamefont {Bowles}, \citenamefont {Puig}, \citenamefont {Recio-Armengol}, \citenamefont {Holmes},\ and\ \citenamefont {Thanasilp}}]{Lerch2026IQPBornInit}%
  \BibitemOpen
  \bibfield  {author} {\bibinfo {author} {\bibfnamefont {S.}~\bibnamefont {Lerch}}, \bibinfo {author} {\bibfnamefont {J.}~\bibnamefont {Bowles}}, \bibinfo {author} {\bibfnamefont {R.}~\bibnamefont {Puig}}, \bibinfo {author} {\bibfnamefont {E.}~\bibnamefont {Recio-Armengol}}, \bibinfo {author} {\bibfnamefont {Z.}~\bibnamefont {Holmes}},\ and\ \bibinfo {author} {\bibfnamefont {S.}~\bibnamefont {Thanasilp}},\ }\href@noop {} {\bibinfo {title} {{IQP} {B}orn machines under data-dependent and agnostic initialization strategies}} (\bibinfo {year} {2026}),\ \Eprint {https://arxiv.org/abs/2603.14576} {arXiv:2603.14576 [quant-ph]} \BibitemShut {NoStop}%
\bibitem [{\citenamefont {Herrero-Gonzalez}\ \emph {et~al.}(2025)\citenamefont {Herrero-Gonzalez}, \citenamefont {Coyle}, \citenamefont {McDowall}, \citenamefont {Grassie}, \citenamefont {Beentjes}, \citenamefont {Khamseh},\ and\ \citenamefont {Kashefi}}]{HerreroGonzalez2025BornUltimatum}%
  \BibitemOpen
  \bibfield  {author} {\bibinfo {author} {\bibfnamefont {M.}~\bibnamefont {Herrero-Gonzalez}}, \bibinfo {author} {\bibfnamefont {B.}~\bibnamefont {Coyle}}, \bibinfo {author} {\bibfnamefont {K.}~\bibnamefont {McDowall}}, \bibinfo {author} {\bibfnamefont {R.}~\bibnamefont {Grassie}}, \bibinfo {author} {\bibfnamefont {S.~V.}\ \bibnamefont {Beentjes}}, \bibinfo {author} {\bibfnamefont {A.}~\bibnamefont {Khamseh}},\ and\ \bibinfo {author} {\bibfnamefont {E.}~\bibnamefont {Kashefi}},\ }\href@noop {} {\bibinfo {title} {The {B}orn ultimatum: Conditions for classical surrogation of quantum generative models with correlators}} (\bibinfo {year} {2025}),\ \Eprint {https://arxiv.org/abs/2511.01845} {arXiv:2511.01845 [quant-ph]} \BibitemShut {NoStop}%
\bibitem [{\citenamefont {Huang}\ \emph {et~al.}(2025)\citenamefont {Huang}, \citenamefont {Broughton}, \citenamefont {Eassa}, \citenamefont {Neven}, \citenamefont {Babbush},\ and\ \citenamefont {McClean}}]{Huang2025GenerativeQuantumAdvantage}%
  \BibitemOpen
  \bibfield  {author} {\bibinfo {author} {\bibfnamefont {H.-Y.}\ \bibnamefont {Huang}}, \bibinfo {author} {\bibfnamefont {M.}~\bibnamefont {Broughton}}, \bibinfo {author} {\bibfnamefont {N.~M.}\ \bibnamefont {Eassa}}, \bibinfo {author} {\bibfnamefont {H.}~\bibnamefont {Neven}}, \bibinfo {author} {\bibfnamefont {R.}~\bibnamefont {Babbush}},\ and\ \bibinfo {author} {\bibfnamefont {J.~R.}\ \bibnamefont {McClean}},\ }\href@noop {} {\bibinfo {title} {Generative quantum advantage for classical and quantum problems}} (\bibinfo {year} {2025}),\ \Eprint {https://arxiv.org/abs/2509.09033} {arXiv:2509.09033 [quant-ph]} \BibitemShut {NoStop}%
\bibitem [{\citenamefont {Gretton}\ \emph {et~al.}(2012)\citenamefont {Gretton}, \citenamefont {Borgwardt}, \citenamefont {Rasch}, \citenamefont {Sch{\"o}lkopf},\ and\ \citenamefont {Smola}}]{Gretton2012MMD}%
  \BibitemOpen
  \bibfield  {author} {\bibinfo {author} {\bibfnamefont {A.}~\bibnamefont {Gretton}}, \bibinfo {author} {\bibfnamefont {K.~M.}\ \bibnamefont {Borgwardt}}, \bibinfo {author} {\bibfnamefont {M.~J.}\ \bibnamefont {Rasch}}, \bibinfo {author} {\bibfnamefont {B.}~\bibnamefont {Sch{\"o}lkopf}},\ and\ \bibinfo {author} {\bibfnamefont {A.}~\bibnamefont {Smola}},\ }\href@noop {} {\bibfield  {journal} {\bibinfo  {journal} {Journal of Machine Learning Research}\ }\textbf {\bibinfo {volume} {13}},\ \bibinfo {pages} {723} (\bibinfo {year} {2012})}\BibitemShut {NoStop}%
\bibitem [{\citenamefont {Demidik}\ \emph {et~al.}(2025)\citenamefont {Demidik}, \citenamefont {T{\"u}ys{\"u}z}, \citenamefont {Grossi},\ and\ \citenamefont {Jansen}}]{Demidik2025SampleBased}%
  \BibitemOpen
  \bibfield  {author} {\bibinfo {author} {\bibfnamefont {M.}~\bibnamefont {Demidik}}, \bibinfo {author} {\bibfnamefont {C.}~\bibnamefont {T{\"u}ys{\"u}z}}, \bibinfo {author} {\bibfnamefont {M.}~\bibnamefont {Grossi}},\ and\ \bibinfo {author} {\bibfnamefont {K.}~\bibnamefont {Jansen}},\ }\href@noop {} {\bibinfo {title} {Sample-based training of quantum generative models}} (\bibinfo {year} {2025}),\ \Eprint {https://arxiv.org/abs/2511.11802} {arXiv:2511.11802 [quant-ph]} \BibitemShut {NoStop}%
\bibitem [{\citenamefont {Braunstein}\ and\ \citenamefont {van Loock}(2005)}]{Braunstein2005CVReview}%
  \BibitemOpen
  \bibfield  {author} {\bibinfo {author} {\bibfnamefont {S.~L.}\ \bibnamefont {Braunstein}}\ and\ \bibinfo {author} {\bibfnamefont {P.}~\bibnamefont {van Loock}},\ }\href {https://doi.org/10.1103/RevModPhys.77.513} {\bibfield  {journal} {\bibinfo  {journal} {Reviews of Modern Physics}\ }\textbf {\bibinfo {volume} {77}},\ \bibinfo {pages} {513} (\bibinfo {year} {2005})}\BibitemShut {NoStop}%
\bibitem [{\citenamefont {Weedbrook}\ \emph {et~al.}(2012)\citenamefont {Weedbrook}, \citenamefont {Pirandola}, \citenamefont {Garc{\'i}a-Patr{\'o}n}, \citenamefont {Cerf}, \citenamefont {Ralph}, \citenamefont {Shapiro},\ and\ \citenamefont {Lloyd}}]{Weedbrook2012CVReview}%
  \BibitemOpen
  \bibfield  {author} {\bibinfo {author} {\bibfnamefont {C.}~\bibnamefont {Weedbrook}}, \bibinfo {author} {\bibfnamefont {S.}~\bibnamefont {Pirandola}}, \bibinfo {author} {\bibfnamefont {R.}~\bibnamefont {Garc{\'i}a-Patr{\'o}n}}, \bibinfo {author} {\bibfnamefont {N.~J.}\ \bibnamefont {Cerf}}, \bibinfo {author} {\bibfnamefont {T.~C.}\ \bibnamefont {Ralph}}, \bibinfo {author} {\bibfnamefont {J.~H.}\ \bibnamefont {Shapiro}},\ and\ \bibinfo {author} {\bibfnamefont {S.}~\bibnamefont {Lloyd}},\ }\href {https://doi.org/10.1103/RevModPhys.84.621} {\bibfield  {journal} {\bibinfo  {journal} {Reviews of Modern Physics}\ }\textbf {\bibinfo {volume} {84}},\ \bibinfo {pages} {621} (\bibinfo {year} {2012})}\BibitemShut {NoStop}%
\bibitem [{\citenamefont {Andersen}\ \emph {et~al.}(2015)\citenamefont {Andersen}, \citenamefont {Neergaard-Nielsen}, \citenamefont {van Loock},\ and\ \citenamefont {Furusawa}}]{Andersen2015HybridCV}%
  \BibitemOpen
  \bibfield  {author} {\bibinfo {author} {\bibfnamefont {U.~L.}\ \bibnamefont {Andersen}}, \bibinfo {author} {\bibfnamefont {J.~S.}\ \bibnamefont {Neergaard-Nielsen}}, \bibinfo {author} {\bibfnamefont {P.}~\bibnamefont {van Loock}},\ and\ \bibinfo {author} {\bibfnamefont {A.}~\bibnamefont {Furusawa}},\ }\href {https://doi.org/10.1038/nphys3410} {\bibfield  {journal} {\bibinfo  {journal} {Nature Physics}\ }\textbf {\bibinfo {volume} {11}},\ \bibinfo {pages} {713} (\bibinfo {year} {2015})}\BibitemShut {NoStop}%
\bibitem [{\citenamefont {Reck}\ \emph {et~al.}(1994)\citenamefont {Reck}, \citenamefont {Zeilinger}, \citenamefont {Bernstein},\ and\ \citenamefont {Bertani}}]{Reck1994}%
  \BibitemOpen
  \bibfield  {author} {\bibinfo {author} {\bibfnamefont {M.}~\bibnamefont {Reck}}, \bibinfo {author} {\bibfnamefont {A.}~\bibnamefont {Zeilinger}}, \bibinfo {author} {\bibfnamefont {H.~J.}\ \bibnamefont {Bernstein}},\ and\ \bibinfo {author} {\bibfnamefont {P.}~\bibnamefont {Bertani}},\ }\href {https://doi.org/10.1103/PhysRevLett.73.58} {\bibfield  {journal} {\bibinfo  {journal} {Physical Review Letters}\ }\textbf {\bibinfo {volume} {73}},\ \bibinfo {pages} {58} (\bibinfo {year} {1994})}\BibitemShut {NoStop}%
\bibitem [{\citenamefont {Clements}\ \emph {et~al.}(2016)\citenamefont {Clements}, \citenamefont {Humphreys}, \citenamefont {Metcalf}, \citenamefont {Kolthammer},\ and\ \citenamefont {Walmsley}}]{Clements2016Clements}%
  \BibitemOpen
  \bibfield  {author} {\bibinfo {author} {\bibfnamefont {W.~R.}\ \bibnamefont {Clements}}, \bibinfo {author} {\bibfnamefont {P.~C.}\ \bibnamefont {Humphreys}}, \bibinfo {author} {\bibfnamefont {B.~J.}\ \bibnamefont {Metcalf}}, \bibinfo {author} {\bibfnamefont {W.~S.}\ \bibnamefont {Kolthammer}},\ and\ \bibinfo {author} {\bibfnamefont {I.~A.}\ \bibnamefont {Walmsley}},\ }\href {https://doi.org/10.1364/OPTICA.3.001460} {\bibfield  {journal} {\bibinfo  {journal} {Optica}\ }\textbf {\bibinfo {volume} {3}},\ \bibinfo {pages} {1460} (\bibinfo {year} {2016})}\BibitemShut {NoStop}%
\bibitem [{\citenamefont {Arrazola}\ \emph {et~al.}(2021)\citenamefont {Arrazola} \emph {et~al.}}]{Arrazola2021PhotonicML}%
  \BibitemOpen
  \bibfield  {author} {\bibinfo {author} {\bibfnamefont {J.~M.}\ \bibnamefont {Arrazola}} \emph {et~al.},\ }\href {https://doi.org/10.1038/s41586-021-03202-1} {\bibfield  {journal} {\bibinfo  {journal} {Nature}\ }\textbf {\bibinfo {volume} {591}},\ \bibinfo {pages} {54} (\bibinfo {year} {2021})}\BibitemShut {NoStop}%
\bibitem [{\citenamefont {Hadfield}(2009)}]{Hadfield2009SinglePhotonDetectors}%
  \BibitemOpen
  \bibfield  {author} {\bibinfo {author} {\bibfnamefont {R.~H.}\ \bibnamefont {Hadfield}},\ }\href {https://doi.org/10.1038/nphoton.2009.230} {\bibfield  {journal} {\bibinfo  {journal} {Nature Photonics}\ }\textbf {\bibinfo {volume} {3}},\ \bibinfo {pages} {696} (\bibinfo {year} {2009})}\BibitemShut {NoStop}%
\bibitem [{\citenamefont {Lita}\ \emph {et~al.}(2008)\citenamefont {Lita}, \citenamefont {Miller},\ and\ \citenamefont {Nam}}]{Lita2008TES}%
  \BibitemOpen
  \bibfield  {author} {\bibinfo {author} {\bibfnamefont {A.~E.}\ \bibnamefont {Lita}}, \bibinfo {author} {\bibfnamefont {A.~J.}\ \bibnamefont {Miller}},\ and\ \bibinfo {author} {\bibfnamefont {S.~W.}\ \bibnamefont {Nam}},\ }\href {https://doi.org/10.1364/OE.16.003032} {\bibfield  {journal} {\bibinfo  {journal} {Optics Express}\ }\textbf {\bibinfo {volume} {16}},\ \bibinfo {pages} {3032} (\bibinfo {year} {2008})}\BibitemShut {NoStop}%
\bibitem [{\citenamefont {Lloyd}\ and\ \citenamefont {Braunstein}(1999)}]{LloydBraunstein1999}%
  \BibitemOpen
  \bibfield  {author} {\bibinfo {author} {\bibfnamefont {S.}~\bibnamefont {Lloyd}}\ and\ \bibinfo {author} {\bibfnamefont {S.~L.}\ \bibnamefont {Braunstein}},\ }\href {https://doi.org/10.1103/PhysRevLett.82.1784} {\bibfield  {journal} {\bibinfo  {journal} {Physical Review Letters}\ }\textbf {\bibinfo {volume} {82}},\ \bibinfo {pages} {1784} (\bibinfo {year} {1999})}\BibitemShut {NoStop}%
\bibitem [{\citenamefont {Walschaers}(2021)}]{Walschaers2021NonGaussian}%
  \BibitemOpen
  \bibfield  {author} {\bibinfo {author} {\bibfnamefont {M.}~\bibnamefont {Walschaers}},\ }\href {https://doi.org/10.1103/PRXQuantum.2.030204} {\bibfield  {journal} {\bibinfo  {journal} {PRX Quantum}\ }\textbf {\bibinfo {volume} {2}},\ \bibinfo {pages} {030204} (\bibinfo {year} {2021})}\BibitemShut {NoStop}%
\bibitem [{\citenamefont {Chabaud}\ and\ \citenamefont {Walschaers}(2023)}]{Chabaud2023Resources}%
  \BibitemOpen
  \bibfield  {author} {\bibinfo {author} {\bibfnamefont {U.}~\bibnamefont {Chabaud}}\ and\ \bibinfo {author} {\bibfnamefont {M.}~\bibnamefont {Walschaers}},\ }\href {https://doi.org/10.1103/PhysRevLett.130.090602} {\bibfield  {journal} {\bibinfo  {journal} {Physical Review Letters}\ }\textbf {\bibinfo {volume} {130}},\ \bibinfo {pages} {090602} (\bibinfo {year} {2023})}\BibitemShut {NoStop}%
\bibitem [{\citenamefont {Lund}\ \emph {et~al.}(2014)\citenamefont {Lund}, \citenamefont {Laing}, \citenamefont {Rahimi-Keshari}, \citenamefont {Rudolph}, \citenamefont {O'Brien},\ and\ \citenamefont {Ralph}}]{Lund2014GaussianBS}%
  \BibitemOpen
  \bibfield  {author} {\bibinfo {author} {\bibfnamefont {A.~P.}\ \bibnamefont {Lund}}, \bibinfo {author} {\bibfnamefont {A.}~\bibnamefont {Laing}}, \bibinfo {author} {\bibfnamefont {S.}~\bibnamefont {Rahimi-Keshari}}, \bibinfo {author} {\bibfnamefont {T.}~\bibnamefont {Rudolph}}, \bibinfo {author} {\bibfnamefont {J.~L.}\ \bibnamefont {O'Brien}},\ and\ \bibinfo {author} {\bibfnamefont {T.~C.}\ \bibnamefont {Ralph}},\ }\href {https://doi.org/10.1103/PhysRevLett.113.100502} {\bibfield  {journal} {\bibinfo  {journal} {Physical Review Letters}\ }\textbf {\bibinfo {volume} {113}},\ \bibinfo {pages} {100502} (\bibinfo {year} {2014})}\BibitemShut {NoStop}%
\bibitem [{\citenamefont {Zhong}\ \emph {et~al.}(2020)\citenamefont {Zhong}, \citenamefont {Wang}, \citenamefont {Deng}, \citenamefont {Chen}, \citenamefont {Peng}, \citenamefont {Luo}, \citenamefont {Qin}, \citenamefont {Wu}, \citenamefont {Ding}, \citenamefont {Hu}, \citenamefont {Hu}, \citenamefont {Yang}, \citenamefont {Zhang}, \citenamefont {Li}, \citenamefont {Li}, \citenamefont {Jiang}, \citenamefont {Gan}, \citenamefont {Yang}, \citenamefont {You}, \citenamefont {Wang}, \citenamefont {Li}, \citenamefont {Liu}, \citenamefont {Lu},\ and\ \citenamefont {Pan}}]{Zhong2020Jiuzhang}%
  \BibitemOpen
  \bibfield  {author} {\bibinfo {author} {\bibfnamefont {H.-S.}\ \bibnamefont {Zhong}}, \bibinfo {author} {\bibfnamefont {H.}~\bibnamefont {Wang}}, \bibinfo {author} {\bibfnamefont {Y.-H.}\ \bibnamefont {Deng}}, \bibinfo {author} {\bibfnamefont {M.-C.}\ \bibnamefont {Chen}}, \bibinfo {author} {\bibfnamefont {L.-C.}\ \bibnamefont {Peng}}, \bibinfo {author} {\bibfnamefont {Y.-H.}\ \bibnamefont {Luo}}, \bibinfo {author} {\bibfnamefont {J.}~\bibnamefont {Qin}}, \bibinfo {author} {\bibfnamefont {D.}~\bibnamefont {Wu}}, \bibinfo {author} {\bibfnamefont {X.}~\bibnamefont {Ding}}, \bibinfo {author} {\bibfnamefont {Y.}~\bibnamefont {Hu}}, \bibinfo {author} {\bibfnamefont {P.}~\bibnamefont {Hu}}, \bibinfo {author} {\bibfnamefont {X.-Y.}\ \bibnamefont {Yang}}, \bibinfo {author} {\bibfnamefont {W.-J.}\ \bibnamefont {Zhang}}, \bibinfo {author} {\bibfnamefont {H.}~\bibnamefont {Li}}, \bibinfo {author} {\bibfnamefont {Y.}~\bibnamefont {Li}}, \bibinfo {author} {\bibfnamefont {X.}~\bibnamefont {Jiang}}, \bibinfo {author}
  {\bibfnamefont {L.}~\bibnamefont {Gan}}, \bibinfo {author} {\bibfnamefont {G.}~\bibnamefont {Yang}}, \bibinfo {author} {\bibfnamefont {L.}~\bibnamefont {You}}, \bibinfo {author} {\bibfnamefont {Z.}~\bibnamefont {Wang}}, \bibinfo {author} {\bibfnamefont {L.}~\bibnamefont {Li}}, \bibinfo {author} {\bibfnamefont {N.-L.}\ \bibnamefont {Liu}}, \bibinfo {author} {\bibfnamefont {C.-Y.}\ \bibnamefont {Lu}},\ and\ \bibinfo {author} {\bibfnamefont {J.-W.}\ \bibnamefont {Pan}},\ }\href {https://doi.org/10.1126/science.abe8770} {\bibfield  {journal} {\bibinfo  {journal} {Science}\ }\textbf {\bibinfo {volume} {370}},\ \bibinfo {pages} {1460} (\bibinfo {year} {2020})}\BibitemShut {NoStop}%
\bibitem [{\citenamefont {Madsen}\ \emph {et~al.}(2022)\citenamefont {Madsen}, \citenamefont {Laudenbach}, \citenamefont {Askarani}, \citenamefont {Rortais}, \citenamefont {Vincent}, \citenamefont {Bulmer}, \citenamefont {Miatto}, \citenamefont {Neuhaus}, \citenamefont {Helt}, \citenamefont {Collins}, \citenamefont {Lita}, \citenamefont {Gerrits}, \citenamefont {Nam}, \citenamefont {Vaidya}, \citenamefont {Menotti}, \citenamefont {Dhand}, \citenamefont {Vernon}, \citenamefont {Quesada},\ and\ \citenamefont {Lavoie}}]{Madsen2022PhotonicAdvantage}%
  \BibitemOpen
  \bibfield  {author} {\bibinfo {author} {\bibfnamefont {L.~S.}\ \bibnamefont {Madsen}}, \bibinfo {author} {\bibfnamefont {F.}~\bibnamefont {Laudenbach}}, \bibinfo {author} {\bibfnamefont {M.~F.}\ \bibnamefont {Askarani}}, \bibinfo {author} {\bibfnamefont {F.}~\bibnamefont {Rortais}}, \bibinfo {author} {\bibfnamefont {T.}~\bibnamefont {Vincent}}, \bibinfo {author} {\bibfnamefont {J.~F.~F.}\ \bibnamefont {Bulmer}}, \bibinfo {author} {\bibfnamefont {F.~M.}\ \bibnamefont {Miatto}}, \bibinfo {author} {\bibfnamefont {L.}~\bibnamefont {Neuhaus}}, \bibinfo {author} {\bibfnamefont {L.~G.}\ \bibnamefont {Helt}}, \bibinfo {author} {\bibfnamefont {M.~J.}\ \bibnamefont {Collins}}, \bibinfo {author} {\bibfnamefont {A.~E.}\ \bibnamefont {Lita}}, \bibinfo {author} {\bibfnamefont {T.}~\bibnamefont {Gerrits}}, \bibinfo {author} {\bibfnamefont {S.~W.}\ \bibnamefont {Nam}}, \bibinfo {author} {\bibfnamefont {V.~D.}\ \bibnamefont {Vaidya}}, \bibinfo {author} {\bibfnamefont {M.}~\bibnamefont {Menotti}}, \bibinfo {author}
  {\bibfnamefont {I.}~\bibnamefont {Dhand}}, \bibinfo {author} {\bibfnamefont {Z.}~\bibnamefont {Vernon}}, \bibinfo {author} {\bibfnamefont {N.}~\bibnamefont {Quesada}},\ and\ \bibinfo {author} {\bibfnamefont {J.}~\bibnamefont {Lavoie}},\ }\href {https://doi.org/10.1038/s41586-022-04725-x} {\bibfield  {journal} {\bibinfo  {journal} {Nature}\ }\textbf {\bibinfo {volume} {606}},\ \bibinfo {pages} {75} (\bibinfo {year} {2022})}\BibitemShut {NoStop}%
\bibitem [{\citenamefont {Gottesman}\ \emph {et~al.}(2001)\citenamefont {Gottesman}, \citenamefont {Kitaev},\ and\ \citenamefont {Preskill}}]{Gottesman2001GKP}%
  \BibitemOpen
  \bibfield  {author} {\bibinfo {author} {\bibfnamefont {D.}~\bibnamefont {Gottesman}}, \bibinfo {author} {\bibfnamefont {A.}~\bibnamefont {Kitaev}},\ and\ \bibinfo {author} {\bibfnamefont {J.}~\bibnamefont {Preskill}},\ }\href {https://doi.org/10.1103/PhysRevA.64.012310} {\bibfield  {journal} {\bibinfo  {journal} {Physical Review A}\ }\textbf {\bibinfo {volume} {64}},\ \bibinfo {pages} {012310} (\bibinfo {year} {2001})}\BibitemShut {NoStop}%
\bibitem [{\citenamefont {Konno}\ \emph {et~al.}(2024)\citenamefont {Konno}, \citenamefont {Asavanant}, \citenamefont {Hanamura}, \citenamefont {Nagayoshi}, \citenamefont {Fukui}, \citenamefont {Sakaguchi}, \citenamefont {Ide}, \citenamefont {China}, \citenamefont {Yabuno}, \citenamefont {Miki}, \citenamefont {Terai}, \citenamefont {Takase}, \citenamefont {Endo}, \citenamefont {Marek}, \citenamefont {Filip}, \citenamefont {van Loock},\ and\ \citenamefont {Furusawa}}]{Konno2024LogicalStates}%
  \BibitemOpen
  \bibfield  {author} {\bibinfo {author} {\bibfnamefont {S.}~\bibnamefont {Konno}}, \bibinfo {author} {\bibfnamefont {W.}~\bibnamefont {Asavanant}}, \bibinfo {author} {\bibfnamefont {F.}~\bibnamefont {Hanamura}}, \bibinfo {author} {\bibfnamefont {H.}~\bibnamefont {Nagayoshi}}, \bibinfo {author} {\bibfnamefont {K.}~\bibnamefont {Fukui}}, \bibinfo {author} {\bibfnamefont {A.}~\bibnamefont {Sakaguchi}}, \bibinfo {author} {\bibfnamefont {R.}~\bibnamefont {Ide}}, \bibinfo {author} {\bibfnamefont {F.}~\bibnamefont {China}}, \bibinfo {author} {\bibfnamefont {M.}~\bibnamefont {Yabuno}}, \bibinfo {author} {\bibfnamefont {S.}~\bibnamefont {Miki}}, \bibinfo {author} {\bibfnamefont {H.}~\bibnamefont {Terai}}, \bibinfo {author} {\bibfnamefont {K.}~\bibnamefont {Takase}}, \bibinfo {author} {\bibfnamefont {M.}~\bibnamefont {Endo}}, \bibinfo {author} {\bibfnamefont {P.}~\bibnamefont {Marek}}, \bibinfo {author} {\bibfnamefont {R.}~\bibnamefont {Filip}}, \bibinfo {author} {\bibfnamefont {P.}~\bibnamefont {van Loock}},\ and\
  \bibinfo {author} {\bibfnamefont {A.}~\bibnamefont {Furusawa}},\ }\href {https://doi.org/10.1126/science.adk7560} {\bibfield  {journal} {\bibinfo  {journal} {Science}\ }\textbf {\bibinfo {volume} {383}},\ \bibinfo {pages} {289} (\bibinfo {year} {2024})}\BibitemShut {NoStop}%
\bibitem [{\citenamefont {Asavanant}\ \emph {et~al.}(2019)\citenamefont {Asavanant}, \citenamefont {Shiozawa}, \citenamefont {Yokoyama}, \citenamefont {Charoensombutamon}, \citenamefont {Emura}, \citenamefont {Alexander}, \citenamefont {Takeda}, \citenamefont {Yoshikawa}, \citenamefont {Menicucci}, \citenamefont {Yonezawa},\ and\ \citenamefont {Furusawa}}]{Asavanant2019}%
  \BibitemOpen
  \bibfield  {author} {\bibinfo {author} {\bibfnamefont {W.}~\bibnamefont {Asavanant}}, \bibinfo {author} {\bibfnamefont {Y.}~\bibnamefont {Shiozawa}}, \bibinfo {author} {\bibfnamefont {S.}~\bibnamefont {Yokoyama}}, \bibinfo {author} {\bibfnamefont {B.}~\bibnamefont {Charoensombutamon}}, \bibinfo {author} {\bibfnamefont {H.}~\bibnamefont {Emura}}, \bibinfo {author} {\bibfnamefont {R.~N.}\ \bibnamefont {Alexander}}, \bibinfo {author} {\bibfnamefont {S.}~\bibnamefont {Takeda}}, \bibinfo {author} {\bibfnamefont {J.-i.}\ \bibnamefont {Yoshikawa}}, \bibinfo {author} {\bibfnamefont {N.~C.}\ \bibnamefont {Menicucci}}, \bibinfo {author} {\bibfnamefont {H.}~\bibnamefont {Yonezawa}},\ and\ \bibinfo {author} {\bibfnamefont {A.}~\bibnamefont {Furusawa}},\ }\href {https://doi.org/10.1126/science.aay2645} {\bibfield  {journal} {\bibinfo  {journal} {Science}\ }\textbf {\bibinfo {volume} {366}},\ \bibinfo {pages} {373} (\bibinfo {year} {2019})}\BibitemShut {NoStop}%
\bibitem [{\citenamefont {Larsen}\ \emph {et~al.}(2019)\citenamefont {Larsen}, \citenamefont {Guo}, \citenamefont {Breum}, \citenamefont {Neergaard-Nielsen},\ and\ \citenamefont {Andersen}}]{Larsen2019}%
  \BibitemOpen
  \bibfield  {author} {\bibinfo {author} {\bibfnamefont {M.~V.}\ \bibnamefont {Larsen}}, \bibinfo {author} {\bibfnamefont {X.}~\bibnamefont {Guo}}, \bibinfo {author} {\bibfnamefont {C.~R.}\ \bibnamefont {Breum}}, \bibinfo {author} {\bibfnamefont {J.~S.}\ \bibnamefont {Neergaard-Nielsen}},\ and\ \bibinfo {author} {\bibfnamefont {U.~L.}\ \bibnamefont {Andersen}},\ }\href {https://doi.org/10.1126/science.aay4354} {\bibfield  {journal} {\bibinfo  {journal} {Science}\ }\textbf {\bibinfo {volume} {366}},\ \bibinfo {pages} {369} (\bibinfo {year} {2019})}\BibitemShut {NoStop}%
\bibitem [{\citenamefont {Yoon}\ \emph {et~al.}(2026)\citenamefont {Yoon}, \citenamefont {Roh}, \citenamefont {Gwak},\ and\ \citenamefont {Ra}}]{Yoon2026Homodyne}%
  \BibitemOpen
  \bibfield  {author} {\bibinfo {author} {\bibfnamefont {Y.-D.}\ \bibnamefont {Yoon}}, \bibinfo {author} {\bibfnamefont {C.}~\bibnamefont {Roh}}, \bibinfo {author} {\bibfnamefont {G.}~\bibnamefont {Gwak}},\ and\ \bibinfo {author} {\bibfnamefont {Y.-S.}\ \bibnamefont {Ra}},\ }\href@noop {} {\bibinfo {title} {Camera-enabled scalable homodyne detection of multimode quantum light}} (\bibinfo {year} {2026}),\ \Eprint {https://arxiv.org/abs/2606.10387} {arXiv:2606.10387 [quant-ph]} \BibitemShut {NoStop}%
\bibitem [{\citenamefont {Cimini}\ \emph {et~al.}(2026)\citenamefont {Cimini}, \citenamefont {Sohoni}, \citenamefont {Presutti}, \citenamefont {Malia}, \citenamefont {Ma}, \citenamefont {Yanagimoto}, \citenamefont {Wang}, \citenamefont {Onodera}, \citenamefont {Wright},\ and\ \citenamefont {McMahon}}]{Cimini2026GBSReservoir}%
  \BibitemOpen
  \bibfield  {author} {\bibinfo {author} {\bibfnamefont {V.}~\bibnamefont {Cimini}}, \bibinfo {author} {\bibfnamefont {M.~M.}\ \bibnamefont {Sohoni}}, \bibinfo {author} {\bibfnamefont {F.}~\bibnamefont {Presutti}}, \bibinfo {author} {\bibfnamefont {B.~K.}\ \bibnamefont {Malia}}, \bibinfo {author} {\bibfnamefont {S.-Y.}\ \bibnamefont {Ma}}, \bibinfo {author} {\bibfnamefont {R.}~\bibnamefont {Yanagimoto}}, \bibinfo {author} {\bibfnamefont {T.}~\bibnamefont {Wang}}, \bibinfo {author} {\bibfnamefont {T.}~\bibnamefont {Onodera}}, \bibinfo {author} {\bibfnamefont {L.~G.}\ \bibnamefont {Wright}},\ and\ \bibinfo {author} {\bibfnamefont {P.~L.}\ \bibnamefont {McMahon}},\ }\href {https://doi.org/10.1038/s41534-026-01251-w} {\bibfield  {journal} {\bibinfo  {journal} {npj Quantum Information}\ }\textbf {\bibinfo {volume} {12}},\ \bibinfo {pages} {104} (\bibinfo {year} {2026})}\BibitemShut {NoStop}%
\bibitem [{\citenamefont {Killoran}\ \emph {et~al.}(2019{\natexlab{a}})\citenamefont {Killoran}, \citenamefont {Bromley}, \citenamefont {Arrazola}, \citenamefont {Schuld}, \citenamefont {Quesada},\ and\ \citenamefont {Lloyd}}]{Killoran2019CVQNN}%
  \BibitemOpen
  \bibfield  {author} {\bibinfo {author} {\bibfnamefont {N.}~\bibnamefont {Killoran}}, \bibinfo {author} {\bibfnamefont {T.~R.}\ \bibnamefont {Bromley}}, \bibinfo {author} {\bibfnamefont {J.~M.}\ \bibnamefont {Arrazola}}, \bibinfo {author} {\bibfnamefont {M.}~\bibnamefont {Schuld}}, \bibinfo {author} {\bibfnamefont {N.}~\bibnamefont {Quesada}},\ and\ \bibinfo {author} {\bibfnamefont {S.}~\bibnamefont {Lloyd}},\ }\href {https://doi.org/10.1103/PhysRevResearch.1.033063} {\bibfield  {journal} {\bibinfo  {journal} {Physical Review Research}\ }\textbf {\bibinfo {volume} {1}},\ \bibinfo {pages} {033063} (\bibinfo {year} {2019}{\natexlab{a}})}\BibitemShut {NoStop}%
\bibitem [{\citenamefont {Steinbrecher}\ \emph {et~al.}(2019)\citenamefont {Steinbrecher}, \citenamefont {Olson}, \citenamefont {Englund},\ and\ \citenamefont {Carolan}}]{Steinbrecher2019QONN}%
  \BibitemOpen
  \bibfield  {author} {\bibinfo {author} {\bibfnamefont {G.~R.}\ \bibnamefont {Steinbrecher}}, \bibinfo {author} {\bibfnamefont {J.~P.}\ \bibnamefont {Olson}}, \bibinfo {author} {\bibfnamefont {D.}~\bibnamefont {Englund}},\ and\ \bibinfo {author} {\bibfnamefont {J.}~\bibnamefont {Carolan}},\ }\href {https://doi.org/10.1038/s41534-019-0174-7} {\bibfield  {journal} {\bibinfo  {journal} {npj Quantum Information}\ }\textbf {\bibinfo {volume} {5}},\ \bibinfo {pages} {60} (\bibinfo {year} {2019})}\BibitemShut {NoStop}%
\bibitem [{\citenamefont {Liu}\ \emph {et~al.}(2025)\citenamefont {Liu}, \citenamefont {Brunel}, \citenamefont {Østergaard}, \citenamefont {Cordero}, \citenamefont {Chen}, \citenamefont {Wong}, \citenamefont {Nielsen}, \citenamefont {Bregnsbo}, \citenamefont {Zhou}, \citenamefont {Huang}, \citenamefont {Oh}, \citenamefont {Jiang}, \citenamefont {Preskill}, \citenamefont {Neergaard-Nielsen},\ and\ \citenamefont {Andersen}}]{Andersen2025}%
  \BibitemOpen
  \bibfield  {author} {\bibinfo {author} {\bibfnamefont {Z.-H.}\ \bibnamefont {Liu}}, \bibinfo {author} {\bibfnamefont {R.}~\bibnamefont {Brunel}}, \bibinfo {author} {\bibfnamefont {E.~E.~B.}\ \bibnamefont {Østergaard}}, \bibinfo {author} {\bibfnamefont {O.}~\bibnamefont {Cordero}}, \bibinfo {author} {\bibfnamefont {S.}~\bibnamefont {Chen}}, \bibinfo {author} {\bibfnamefont {Y.}~\bibnamefont {Wong}}, \bibinfo {author} {\bibfnamefont {J.~A.~H.}\ \bibnamefont {Nielsen}}, \bibinfo {author} {\bibfnamefont {A.~B.}\ \bibnamefont {Bregnsbo}}, \bibinfo {author} {\bibfnamefont {S.}~\bibnamefont {Zhou}}, \bibinfo {author} {\bibfnamefont {H.-Y.}\ \bibnamefont {Huang}}, \bibinfo {author} {\bibfnamefont {C.}~\bibnamefont {Oh}}, \bibinfo {author} {\bibfnamefont {L.}~\bibnamefont {Jiang}}, \bibinfo {author} {\bibfnamefont {J.}~\bibnamefont {Preskill}}, \bibinfo {author} {\bibfnamefont {J.~S.}\ \bibnamefont {Neergaard-Nielsen}},\ and\ \bibinfo {author} {\bibfnamefont {U.~L.}\ \bibnamefont {Andersen}},\ }\href
  {https://doi.org/10.1126/science.adv2560} {\bibfield  {journal} {\bibinfo  {journal} {Science}\ }\textbf {\bibinfo {volume} {389}},\ \bibinfo {pages} {1332} (\bibinfo {year} {2025})},\ \Eprint {https://arxiv.org/abs/https://www.science.org/doi/pdf/10.1126/science.adv2560} {https://www.science.org/doi/pdf/10.1126/science.adv2560} \BibitemShut {NoStop}%
\bibitem [{\citenamefont {Jastrzebski}\ \emph {et~al.}(2026)\citenamefont {Jastrzebski}, \citenamefont {Yu}, \citenamefont {Patel},\ and\ \citenamefont {Kyriienko}}]{Jastrzebski2026RelationalLearning}%
  \BibitemOpen
  \bibfield  {author} {\bibinfo {author} {\bibfnamefont {M.}~\bibnamefont {Jastrzebski}}, \bibinfo {author} {\bibfnamefont {S.}~\bibnamefont {Yu}}, \bibinfo {author} {\bibfnamefont {R.~B.}\ \bibnamefont {Patel}},\ and\ \bibinfo {author} {\bibfnamefont {O.}~\bibnamefont {Kyriienko}},\ }\href@noop {} {\bibinfo {title} {Adaptive relational learning on multi-instance quantum data with photonic processors}} (\bibinfo {year} {2026}),\ \Eprint {https://arxiv.org/abs/2609.17352} {arXiv:2609.17352 [quant-ph]} \BibitemShut {NoStop}%
\bibitem [{\citenamefont {Arrazola}\ \emph {et~al.}(2018)\citenamefont {Arrazola}, \citenamefont {Bromley},\ and\ \citenamefont {Rebentrost}}]{Arrazola2018MachineLearningPhotonic}%
  \BibitemOpen
  \bibfield  {author} {\bibinfo {author} {\bibfnamefont {J.~M.}\ \bibnamefont {Arrazola}}, \bibinfo {author} {\bibfnamefont {T.~R.}\ \bibnamefont {Bromley}},\ and\ \bibinfo {author} {\bibfnamefont {P.}~\bibnamefont {Rebentrost}},\ }\href {https://doi.org/10.1103/PhysRevA.98.012322} {\bibfield  {journal} {\bibinfo  {journal} {Physical Review A}\ }\textbf {\bibinfo {volume} {98}},\ \bibinfo {pages} {012322} (\bibinfo {year} {2018})}\BibitemShut {NoStop}%
\bibitem [{\citenamefont {Yu}\ \emph {et~al.}(2023)\citenamefont {Yu}, \citenamefont {Zhong}, \citenamefont {Fang}, \citenamefont {Patel}, \citenamefont {Li}, \citenamefont {Liu}, \citenamefont {Li}, \citenamefont {Xu}, \citenamefont {Sagona-Stophel}, \citenamefont {Mer}, \citenamefont {Thomas}, \citenamefont {Meng}, \citenamefont {Li}, \citenamefont {Yang}, \citenamefont {Wang}, \citenamefont {Guo}, \citenamefont {Zhang}, \citenamefont {Tranmer}, \citenamefont {Dong}, \citenamefont {Wang}, \citenamefont {Tang}, \citenamefont {Li}, \citenamefont {Walmsley},\ and\ \citenamefont {Guo}}]{Yu2023GBSDrugDiscovery}%
  \BibitemOpen
  \bibfield  {author} {\bibinfo {author} {\bibfnamefont {S.}~\bibnamefont {Yu}}, \bibinfo {author} {\bibfnamefont {Z.-P.}\ \bibnamefont {Zhong}}, \bibinfo {author} {\bibfnamefont {Y.}~\bibnamefont {Fang}}, \bibinfo {author} {\bibfnamefont {R.~B.}\ \bibnamefont {Patel}}, \bibinfo {author} {\bibfnamefont {Q.-P.}\ \bibnamefont {Li}}, \bibinfo {author} {\bibfnamefont {W.}~\bibnamefont {Liu}}, \bibinfo {author} {\bibfnamefont {Z.}~\bibnamefont {Li}}, \bibinfo {author} {\bibfnamefont {L.}~\bibnamefont {Xu}}, \bibinfo {author} {\bibfnamefont {S.}~\bibnamefont {Sagona-Stophel}}, \bibinfo {author} {\bibfnamefont {E.}~\bibnamefont {Mer}}, \bibinfo {author} {\bibfnamefont {S.~E.}\ \bibnamefont {Thomas}}, \bibinfo {author} {\bibfnamefont {Y.}~\bibnamefont {Meng}}, \bibinfo {author} {\bibfnamefont {Z.-P.}\ \bibnamefont {Li}}, \bibinfo {author} {\bibfnamefont {Y.-Z.}\ \bibnamefont {Yang}}, \bibinfo {author} {\bibfnamefont {Z.-A.}\ \bibnamefont {Wang}}, \bibinfo {author} {\bibfnamefont {N.-J.}\ \bibnamefont {Guo}}, \bibinfo
  {author} {\bibfnamefont {W.-H.}\ \bibnamefont {Zhang}}, \bibinfo {author} {\bibfnamefont {G.~K.}\ \bibnamefont {Tranmer}}, \bibinfo {author} {\bibfnamefont {Y.}~\bibnamefont {Dong}}, \bibinfo {author} {\bibfnamefont {Y.-T.}\ \bibnamefont {Wang}}, \bibinfo {author} {\bibfnamefont {J.-S.}\ \bibnamefont {Tang}}, \bibinfo {author} {\bibfnamefont {C.-F.}\ \bibnamefont {Li}}, \bibinfo {author} {\bibfnamefont {I.~A.}\ \bibnamefont {Walmsley}},\ and\ \bibinfo {author} {\bibfnamefont {G.-C.}\ \bibnamefont {Guo}},\ }\href {https://doi.org/10.1038/s43588-023-00526-y} {\bibfield  {journal} {\bibinfo  {journal} {Nature Computational Science}\ }\textbf {\bibinfo {volume} {3}},\ \bibinfo {pages} {839} (\bibinfo {year} {2023})}\BibitemShut {NoStop}%
\bibitem [{\citenamefont {Wang}\ and\ \citenamefont {Kyriienko}(2025)}]{Wang2025PhotonicsGCN}%
  \BibitemOpen
  \bibfield  {author} {\bibinfo {author} {\bibfnamefont {Y.}~\bibnamefont {Wang}}\ and\ \bibinfo {author} {\bibfnamefont {O.}~\bibnamefont {Kyriienko}},\ }\href@noop {} {\bibinfo {title} {Photonics-enhanced graph convolutional networks}} (\bibinfo {year} {2025}),\ \Eprint {https://arxiv.org/abs/2512.15549} {arXiv:2512.15549 [quant-ph]} \BibitemShut {NoStop}%
\bibitem [{\citenamefont {Salavrakos}\ \emph {et~al.}(2025)\citenamefont {Salavrakos}, \citenamefont {Sedrakyan}, \citenamefont {Mills}, \citenamefont {Mansfield},\ and\ \citenamefont {Mezher}}]{Salavrakos2024PhotonicBornMachine}%
  \BibitemOpen
  \bibfield  {author} {\bibinfo {author} {\bibfnamefont {A.}~\bibnamefont {Salavrakos}}, \bibinfo {author} {\bibfnamefont {T.}~\bibnamefont {Sedrakyan}}, \bibinfo {author} {\bibfnamefont {J.}~\bibnamefont {Mills}}, \bibinfo {author} {\bibfnamefont {S.}~\bibnamefont {Mansfield}},\ and\ \bibinfo {author} {\bibfnamefont {R.}~\bibnamefont {Mezher}},\ }\href {https://doi.org/10.1103/PhysRevA.111.L030401} {\bibfield  {journal} {\bibinfo  {journal} {Physical Review A}\ }\textbf {\bibinfo {volume} {111}},\ \bibinfo {pages} {L030401} (\bibinfo {year} {2025})},\ \Eprint {https://arxiv.org/abs/2405.02277} {arXiv:2405.02277 [quant-ph]} \BibitemShut {NoStop}%
\bibitem [{\citenamefont {Gottlieb}\ \emph {et~al.}(2026)\citenamefont {Gottlieb}, \citenamefont {Faraji}, \citenamefont {Mezher}, \citenamefont {Ventura}, \citenamefont {Mansfield},\ and\ \citenamefont {Salavrakos}}]{Gottlieb2026PhotonicGenerativeTraining}%
  \BibitemOpen
  \bibfield  {author} {\bibinfo {author} {\bibfnamefont {F.}~\bibnamefont {Gottlieb}}, \bibinfo {author} {\bibfnamefont {C.}~\bibnamefont {Faraji}}, \bibinfo {author} {\bibfnamefont {R.}~\bibnamefont {Mezher}}, \bibinfo {author} {\bibfnamefont {B.}~\bibnamefont {Ventura}}, \bibinfo {author} {\bibfnamefont {S.}~\bibnamefont {Mansfield}},\ and\ \bibinfo {author} {\bibfnamefont {A.}~\bibnamefont {Salavrakos}},\ }\href@noop {} {\bibinfo {title} {Efficient training of photonic quantum generative models}} (\bibinfo {year} {2026}),\ \Eprint {https://arxiv.org/abs/2603.08793} {arXiv:2603.08793 [quant-ph]} \BibitemShut {NoStop}%
\bibitem [{\citenamefont {Sedrakyan}\ and\ \citenamefont {Salavrakos}(2024)}]{Sedrakyan2024PhotonicQGAN}%
  \BibitemOpen
  \bibfield  {author} {\bibinfo {author} {\bibfnamefont {T.}~\bibnamefont {Sedrakyan}}\ and\ \bibinfo {author} {\bibfnamefont {A.}~\bibnamefont {Salavrakos}},\ }\href {https://doi.org/10.1364/opticaq.530346} {\bibfield  {journal} {\bibinfo  {journal} {Optica Quantum}\ }\textbf {\bibinfo {volume} {2}},\ \bibinfo {pages} {458} (\bibinfo {year} {2024})}\BibitemShut {NoStop}%
\bibitem [{\citenamefont {Bacarreza}\ \emph {et~al.}(2025)\citenamefont {Bacarreza}, \citenamefont {Farnsworth}, \citenamefont {Makarovskiy}, \citenamefont {Wallner}, \citenamefont {Hicks}, \citenamefont {Sempere-Llagostera}, \citenamefont {Price}, \citenamefont {Francis-Jones},\ and\ \citenamefont {Clements}}]{Bacarreza2025QuantumLatent}%
  \BibitemOpen
  \bibfield  {author} {\bibinfo {author} {\bibfnamefont {O.}~\bibnamefont {Bacarreza}}, \bibinfo {author} {\bibfnamefont {T.}~\bibnamefont {Farnsworth}}, \bibinfo {author} {\bibfnamefont {A.}~\bibnamefont {Makarovskiy}}, \bibinfo {author} {\bibfnamefont {H.}~\bibnamefont {Wallner}}, \bibinfo {author} {\bibfnamefont {T.}~\bibnamefont {Hicks}}, \bibinfo {author} {\bibfnamefont {S.}~\bibnamefont {Sempere-Llagostera}}, \bibinfo {author} {\bibfnamefont {J.}~\bibnamefont {Price}}, \bibinfo {author} {\bibfnamefont {R.~J.~A.}\ \bibnamefont {Francis-Jones}},\ and\ \bibinfo {author} {\bibfnamefont {W.~R.}\ \bibnamefont {Clements}},\ }\href@noop {} {\bibinfo {title} {Quantum latent distributions in deep generative models}} (\bibinfo {year} {2025}),\ \Eprint {https://arxiv.org/abs/2508.19857} {arXiv:2508.19857 [cs.LG]} \BibitemShut {NoStop}%
\bibitem [{\citenamefont {{\v{C}}epait{\.{e}}}\ \emph {et~al.}(2022)\citenamefont {{\v{C}}epait{\.{e}}}, \citenamefont {Coyle},\ and\ \citenamefont {Kashefi}}]{cepaite2022cvbm}%
  \BibitemOpen
  \bibfield  {author} {\bibinfo {author} {\bibfnamefont {I.}~\bibnamefont {{\v{C}}epait{\.{e}}}}, \bibinfo {author} {\bibfnamefont {B.}~\bibnamefont {Coyle}},\ and\ \bibinfo {author} {\bibfnamefont {E.}~\bibnamefont {Kashefi}},\ }\href {https://doi.org/10.1007/s42484-022-00063-3} {\bibfield  {journal} {\bibinfo  {journal} {Quantum Machine Intelligence}\ }\textbf {\bibinfo {volume} {4}},\ \bibinfo {pages} {6} (\bibinfo {year} {2022})}\BibitemShut {NoStop}%
\bibitem [{\citenamefont {Cont}(2001)}]{Cont2001}%
  \BibitemOpen
  \bibfield  {author} {\bibinfo {author} {\bibfnamefont {R.}~\bibnamefont {Cont}},\ }\href {https://doi.org/10.1080/713665670} {\bibfield  {journal} {\bibinfo  {journal} {Quantitative Finance}\ }\textbf {\bibinfo {volume} {1}},\ \bibinfo {pages} {223} (\bibinfo {year} {2001})}\BibitemShut {NoStop}%
\bibitem [{\citenamefont {Dechant}\ \emph {et~al.}(2026)\citenamefont {Dechant} \emph {et~al.}}]{Dechant2026FinancialQGM}%
  \BibitemOpen
  \bibfield  {author} {\bibinfo {author} {\bibfnamefont {D.}~\bibnamefont {Dechant}} \emph {et~al.},\ }\bibfield  {journal} {\bibinfo  {journal} {Machine Learning: Science and Technology}\ }\href {https://doi.org/10.1088/2632-2153/ae39a2} {10.1088/2632-2153/ae39a2} (\bibinfo {year} {2026})\BibitemShut {NoStop}%
\bibitem [{\citenamefont {Killoran}\ \emph {et~al.}(2019{\natexlab{b}})\citenamefont {Killoran}, \citenamefont {Izaac}, \citenamefont {Quesada}, \citenamefont {Bergholm}, \citenamefont {Amy},\ and\ \citenamefont {Weedbrook}}]{Killoran2019StrawberryFields}%
  \BibitemOpen
  \bibfield  {author} {\bibinfo {author} {\bibfnamefont {N.}~\bibnamefont {Killoran}}, \bibinfo {author} {\bibfnamefont {J.}~\bibnamefont {Izaac}}, \bibinfo {author} {\bibfnamefont {N.}~\bibnamefont {Quesada}}, \bibinfo {author} {\bibfnamefont {V.}~\bibnamefont {Bergholm}}, \bibinfo {author} {\bibfnamefont {M.}~\bibnamefont {Amy}},\ and\ \bibinfo {author} {\bibfnamefont {C.}~\bibnamefont {Weedbrook}},\ }\href {https://doi.org/10.22331/q-2019-03-11-129} {\bibfield  {journal} {\bibinfo  {journal} {Quantum}\ }\textbf {\bibinfo {volume} {3}},\ \bibinfo {pages} {129} (\bibinfo {year} {2019}{\natexlab{b}})}\BibitemShut {NoStop}%
\bibitem [{\citenamefont {Kolarovszki}\ \emph {et~al.}(2025)\citenamefont {Kolarovszki}, \citenamefont {Rybotycki}, \citenamefont {Rakyta}, \citenamefont {Kaposi}, \citenamefont {Po{\'o}r}, \citenamefont {J{\'o}czik}, \citenamefont {Nagy}, \citenamefont {Varga}, \citenamefont {El-Safty}, \citenamefont {Morse}, \citenamefont {Oszmaniec}, \citenamefont {Kozsik},\ and\ \citenamefont {Zimbor{\'a}s}}]{Kolarovszki2024Piquasso}%
  \BibitemOpen
  \bibfield  {author} {\bibinfo {author} {\bibfnamefont {Z.}~\bibnamefont {Kolarovszki}}, \bibinfo {author} {\bibfnamefont {T.}~\bibnamefont {Rybotycki}}, \bibinfo {author} {\bibfnamefont {P.}~\bibnamefont {Rakyta}}, \bibinfo {author} {\bibfnamefont {{\'A}.}~\bibnamefont {Kaposi}}, \bibinfo {author} {\bibfnamefont {B.}~\bibnamefont {Po{\'o}r}}, \bibinfo {author} {\bibfnamefont {S.}~\bibnamefont {J{\'o}czik}}, \bibinfo {author} {\bibfnamefont {D.~T.~R.}\ \bibnamefont {Nagy}}, \bibinfo {author} {\bibfnamefont {H.}~\bibnamefont {Varga}}, \bibinfo {author} {\bibfnamefont {K.~H.}\ \bibnamefont {El-Safty}}, \bibinfo {author} {\bibfnamefont {G.}~\bibnamefont {Morse}}, \bibinfo {author} {\bibfnamefont {M.}~\bibnamefont {Oszmaniec}}, \bibinfo {author} {\bibfnamefont {T.}~\bibnamefont {Kozsik}},\ and\ \bibinfo {author} {\bibfnamefont {Z.}~\bibnamefont {Zimbor{\'a}s}},\ }\href {https://doi.org/10.22331/q-2025-04-15-1708} {\bibfield  {journal} {\bibinfo  {journal} {Quantum}\ }\textbf {\bibinfo {volume} {9}},\ \bibinfo
  {pages} {1708} (\bibinfo {year} {2025})}\BibitemShut {NoStop}%
\bibitem [{\citenamefont {Bartlett}\ \emph {et~al.}(2002)\citenamefont {Bartlett}, \citenamefont {Sanders}, \citenamefont {Braunstein},\ and\ \citenamefont {Nemoto}}]{Bartlett2002Efficient}%
  \BibitemOpen
  \bibfield  {author} {\bibinfo {author} {\bibfnamefont {S.~D.}\ \bibnamefont {Bartlett}}, \bibinfo {author} {\bibfnamefont {B.~C.}\ \bibnamefont {Sanders}}, \bibinfo {author} {\bibfnamefont {S.~L.}\ \bibnamefont {Braunstein}},\ and\ \bibinfo {author} {\bibfnamefont {K.}~\bibnamefont {Nemoto}},\ }\href {https://doi.org/10.1103/PhysRevLett.88.097904} {\bibfield  {journal} {\bibinfo  {journal} {Physical Review Letters}\ }\textbf {\bibinfo {volume} {88}},\ \bibinfo {pages} {097904} (\bibinfo {year} {2002})}\BibitemShut {NoStop}%
\bibitem [{\citenamefont {Mari}\ and\ \citenamefont {Eisert}(2012)}]{MariEisert2012}%
  \BibitemOpen
  \bibfield  {author} {\bibinfo {author} {\bibfnamefont {A.}~\bibnamefont {Mari}}\ and\ \bibinfo {author} {\bibfnamefont {J.}~\bibnamefont {Eisert}},\ }\href {https://doi.org/10.1103/PhysRevLett.109.230503} {\bibfield  {journal} {\bibinfo  {journal} {Physical Review Letters}\ }\textbf {\bibinfo {volume} {109}},\ \bibinfo {pages} {230503} (\bibinfo {year} {2012})}\BibitemShut {NoStop}%
\bibitem [{\citenamefont {Milburn}(1989)}]{Milburn1989KerrQC}%
  \BibitemOpen
  \bibfield  {author} {\bibinfo {author} {\bibfnamefont {G.~J.}\ \bibnamefont {Milburn}},\ }\href {https://doi.org/10.1103/PhysRevLett.62.2124} {\bibfield  {journal} {\bibinfo  {journal} {Physical Review Letters}\ }\textbf {\bibinfo {volume} {62}},\ \bibinfo {pages} {2124} (\bibinfo {year} {1989})}\BibitemShut {NoStop}%
\bibitem [{\citenamefont {Yurke}\ and\ \citenamefont {Stoler}(1986)}]{Yurke1986}%
  \BibitemOpen
  \bibfield  {author} {\bibinfo {author} {\bibfnamefont {B.}~\bibnamefont {Yurke}}\ and\ \bibinfo {author} {\bibfnamefont {D.}~\bibnamefont {Stoler}},\ }\href {https://doi.org/10.1103/PhysRevLett.57.13} {\bibfield  {journal} {\bibinfo  {journal} {Physical Review Letters}\ }\textbf {\bibinfo {volume} {57}},\ \bibinfo {pages} {13} (\bibinfo {year} {1986})}\BibitemShut {NoStop}%
\bibitem [{\citenamefont {Hubbard}(1959)}]{Hubbard1959}%
  \BibitemOpen
  \bibfield  {author} {\bibinfo {author} {\bibfnamefont {J.}~\bibnamefont {Hubbard}},\ }\href {https://doi.org/10.1103/PhysRevLett.3.77} {\bibfield  {journal} {\bibinfo  {journal} {Physical Review Letters}\ }\textbf {\bibinfo {volume} {3}},\ \bibinfo {pages} {77} (\bibinfo {year} {1959})}\BibitemShut {NoStop}%
\bibitem [{\citenamefont {Stratonovich}(1957)}]{Stratonovich1957}%
  \BibitemOpen
  \bibfield  {author} {\bibinfo {author} {\bibfnamefont {R.~L.}\ \bibnamefont {Stratonovich}},\ }\href@noop {} {\bibfield  {journal} {\bibinfo  {journal} {Soviet Physics Doklady}\ }\textbf {\bibinfo {volume} {2}},\ \bibinfo {pages} {416} (\bibinfo {year} {1957})}\BibitemShut {NoStop}%
\bibitem [{\citenamefont {Levin}(1996)}]{Levin1996Oscillatory}%
  \BibitemOpen
  \bibfield  {author} {\bibinfo {author} {\bibfnamefont {D.}~\bibnamefont {Levin}},\ }\href {https://doi.org/10.1016/0377-0427(94)00118-9} {\bibfield  {journal} {\bibinfo  {journal} {Journal of Computational and Applied Mathematics}\ }\textbf {\bibinfo {volume} {67}},\ \bibinfo {pages} {95} (\bibinfo {year} {1996})}\BibitemShut {NoStop}%
\bibitem [{\citenamefont {Byrd}\ \emph {et~al.}(1995)\citenamefont {Byrd}, \citenamefont {Lu}, \citenamefont {Nocedal},\ and\ \citenamefont {Zhu}}]{Byrd1995LBFGSB}%
  \BibitemOpen
  \bibfield  {author} {\bibinfo {author} {\bibfnamefont {R.~H.}\ \bibnamefont {Byrd}}, \bibinfo {author} {\bibfnamefont {P.}~\bibnamefont {Lu}}, \bibinfo {author} {\bibfnamefont {J.}~\bibnamefont {Nocedal}},\ and\ \bibinfo {author} {\bibfnamefont {C.}~\bibnamefont {Zhu}},\ }\href {https://doi.org/10.1137/0916069} {\bibfield  {journal} {\bibinfo  {journal} {SIAM Journal on Scientific Computing}\ }\textbf {\bibinfo {volume} {16}},\ \bibinfo {pages} {1190} (\bibinfo {year} {1995})}\BibitemShut {NoStop}%
\bibitem [{\citenamefont {Aroussi}(2023)}]{Aroussi2023yfinance}%
  \BibitemOpen
  \bibfield  {author} {\bibinfo {author} {\bibfnamefont {R.}~\bibnamefont {Aroussi}},\ }\href {https://github.com/ranaroussi/yfinance} {\bibinfo {title} {yfinance: Download market data from {Y}ahoo! {F}inance's {API}}},\ \bibinfo {howpublished} {GitHub repository} (\bibinfo {year} {2023})\BibitemShut {NoStop}%
\bibitem [{\citenamefont {Spagnolo}\ \emph {et~al.}(2023)\citenamefont {Spagnolo}, \citenamefont {Brod}, \citenamefont {Galv{\~a}o},\ and\ \citenamefont {Sciarrino}}]{Spagnolo2023NonlinearBS}%
  \BibitemOpen
  \bibfield  {author} {\bibinfo {author} {\bibfnamefont {N.}~\bibnamefont {Spagnolo}}, \bibinfo {author} {\bibfnamefont {D.~J.}\ \bibnamefont {Brod}}, \bibinfo {author} {\bibfnamefont {E.~F.}\ \bibnamefont {Galv{\~a}o}},\ and\ \bibinfo {author} {\bibfnamefont {F.}~\bibnamefont {Sciarrino}},\ }\href {https://doi.org/10.1038/s41534-023-00676-x} {\bibfield  {journal} {\bibinfo  {journal} {npj Quantum Information}\ }\textbf {\bibinfo {volume} {9}},\ \bibinfo {pages} {3} (\bibinfo {year} {2023})}\BibitemShut {NoStop}%
\bibitem [{\citenamefont {Douce}\ \emph {et~al.}(2017)\citenamefont {Douce}, \citenamefont {Markham}, \citenamefont {Kashefi}, \citenamefont {Diamanti}, \citenamefont {Coudreau}, \citenamefont {Milman}, \citenamefont {van Loock},\ and\ \citenamefont {Ferrini}}]{Douce2017CVIQP}%
  \BibitemOpen
  \bibfield  {author} {\bibinfo {author} {\bibfnamefont {T.}~\bibnamefont {Douce}}, \bibinfo {author} {\bibfnamefont {D.}~\bibnamefont {Markham}}, \bibinfo {author} {\bibfnamefont {E.}~\bibnamefont {Kashefi}}, \bibinfo {author} {\bibfnamefont {E.}~\bibnamefont {Diamanti}}, \bibinfo {author} {\bibfnamefont {T.}~\bibnamefont {Coudreau}}, \bibinfo {author} {\bibfnamefont {P.}~\bibnamefont {Milman}}, \bibinfo {author} {\bibfnamefont {P.}~\bibnamefont {van Loock}},\ and\ \bibinfo {author} {\bibfnamefont {G.}~\bibnamefont {Ferrini}},\ }\href {https://doi.org/10.1103/PhysRevLett.118.070503} {\bibfield  {journal} {\bibinfo  {journal} {Physical Review Letters}\ }\textbf {\bibinfo {volume} {118}},\ \bibinfo {pages} {070503} (\bibinfo {year} {2017})}\BibitemShut {NoStop}%
\bibitem [{\citenamefont {Chabaud}\ \emph {et~al.}(2017)\citenamefont {Chabaud}, \citenamefont {Douce}, \citenamefont {Markham}, \citenamefont {van Loock}, \citenamefont {Kashefi},\ and\ \citenamefont {Ferrini}}]{Chabaud2017CVSampling}%
  \BibitemOpen
  \bibfield  {author} {\bibinfo {author} {\bibfnamefont {U.}~\bibnamefont {Chabaud}}, \bibinfo {author} {\bibfnamefont {T.}~\bibnamefont {Douce}}, \bibinfo {author} {\bibfnamefont {D.}~\bibnamefont {Markham}}, \bibinfo {author} {\bibfnamefont {P.}~\bibnamefont {van Loock}}, \bibinfo {author} {\bibfnamefont {E.}~\bibnamefont {Kashefi}},\ and\ \bibinfo {author} {\bibfnamefont {G.}~\bibnamefont {Ferrini}},\ }\href {https://doi.org/10.1103/PhysRevA.96.062307} {\bibfield  {journal} {\bibinfo  {journal} {Physical Review A}\ }\textbf {\bibinfo {volume} {96}},\ \bibinfo {pages} {062307} (\bibinfo {year} {2017})}\BibitemShut {NoStop}%
\bibitem [{\citenamefont {Flamini}\ \emph {et~al.}(2019)\citenamefont {Flamini}, \citenamefont {Spagnolo},\ and\ \citenamefont {Sciarrino}}]{Flamini2019PhotonicQML}%
  \BibitemOpen
  \bibfield  {author} {\bibinfo {author} {\bibfnamefont {F.}~\bibnamefont {Flamini}}, \bibinfo {author} {\bibfnamefont {N.}~\bibnamefont {Spagnolo}},\ and\ \bibinfo {author} {\bibfnamefont {F.}~\bibnamefont {Sciarrino}},\ }\href {https://doi.org/10.1088/1361-6633/aad5b2} {\bibfield  {journal} {\bibinfo  {journal} {Reports on Progress in Physics}\ }\textbf {\bibinfo {volume} {82}},\ \bibinfo {pages} {016001} (\bibinfo {year} {2019})}\BibitemShut {NoStop}%
\bibitem [{\citenamefont {Niset}\ \emph {et~al.}(2009)\citenamefont {Niset}, \citenamefont {Fiur{\'a}{\v{s}}ek},\ and\ \citenamefont {Cerf}}]{Niset2009NoGo}%
  \BibitemOpen
  \bibfield  {author} {\bibinfo {author} {\bibfnamefont {J.}~\bibnamefont {Niset}}, \bibinfo {author} {\bibfnamefont {J.}~\bibnamefont {Fiur{\'a}{\v{s}}ek}},\ and\ \bibinfo {author} {\bibfnamefont {N.~J.}\ \bibnamefont {Cerf}},\ }\href {https://doi.org/10.1103/PhysRevLett.102.120501} {\bibfield  {journal} {\bibinfo  {journal} {Physical Review Letters}\ }\textbf {\bibinfo {volume} {102}},\ \bibinfo {pages} {120501} (\bibinfo {year} {2009})}\BibitemShut {NoStop}%
\bibitem [{\citenamefont {Vahlbruch}\ \emph {et~al.}(2016)\citenamefont {Vahlbruch}, \citenamefont {Mehmet}, \citenamefont {Danzmann},\ and\ \citenamefont {Schnabel}}]{Vahlbruch2016Squeezing15dB}%
  \BibitemOpen
  \bibfield  {author} {\bibinfo {author} {\bibfnamefont {H.}~\bibnamefont {Vahlbruch}}, \bibinfo {author} {\bibfnamefont {M.}~\bibnamefont {Mehmet}}, \bibinfo {author} {\bibfnamefont {K.}~\bibnamefont {Danzmann}},\ and\ \bibinfo {author} {\bibfnamefont {R.}~\bibnamefont {Schnabel}},\ }\href {https://doi.org/10.1103/PhysRevLett.117.110801} {\bibfield  {journal} {\bibinfo  {journal} {Physical Review Letters}\ }\textbf {\bibinfo {volume} {117}},\ \bibinfo {pages} {110801} (\bibinfo {year} {2016})}\BibitemShut {NoStop}%
\bibitem [{\citenamefont {Vaidya}\ \emph {et~al.}(2020)\citenamefont {Vaidya}, \citenamefont {Morrison}, \citenamefont {Helt}, \citenamefont {Shahrokshahi}, \citenamefont {Mahler}, \citenamefont {Collins}, \citenamefont {Tan}, \citenamefont {Lavoie}, \citenamefont {Repingon}, \citenamefont {Menotti}, \citenamefont {Quesada}, \citenamefont {Pooser}, \citenamefont {Lita}, \citenamefont {Gerrits}, \citenamefont {Nam},\ and\ \citenamefont {Vernon}}]{Vaidya2020NanophotonicSqueezing}%
  \BibitemOpen
  \bibfield  {author} {\bibinfo {author} {\bibfnamefont {V.~D.}\ \bibnamefont {Vaidya}}, \bibinfo {author} {\bibfnamefont {B.}~\bibnamefont {Morrison}}, \bibinfo {author} {\bibfnamefont {L.~G.}\ \bibnamefont {Helt}}, \bibinfo {author} {\bibfnamefont {R.}~\bibnamefont {Shahrokshahi}}, \bibinfo {author} {\bibfnamefont {D.~H.}\ \bibnamefont {Mahler}}, \bibinfo {author} {\bibfnamefont {M.~J.}\ \bibnamefont {Collins}}, \bibinfo {author} {\bibfnamefont {K.}~\bibnamefont {Tan}}, \bibinfo {author} {\bibfnamefont {J.}~\bibnamefont {Lavoie}}, \bibinfo {author} {\bibfnamefont {A.}~\bibnamefont {Repingon}}, \bibinfo {author} {\bibfnamefont {M.}~\bibnamefont {Menotti}}, \bibinfo {author} {\bibfnamefont {N.}~\bibnamefont {Quesada}}, \bibinfo {author} {\bibfnamefont {R.~C.}\ \bibnamefont {Pooser}}, \bibinfo {author} {\bibfnamefont {A.~E.}\ \bibnamefont {Lita}}, \bibinfo {author} {\bibfnamefont {T.}~\bibnamefont {Gerrits}}, \bibinfo {author} {\bibfnamefont {S.~W.}\ \bibnamefont {Nam}},\ and\ \bibinfo {author} {\bibfnamefont
  {Z.}~\bibnamefont {Vernon}},\ }\href {https://doi.org/10.1126/sciadv.aba9186} {\bibfield  {journal} {\bibinfo  {journal} {Science Advances}\ }\textbf {\bibinfo {volume} {6}},\ \bibinfo {pages} {eaba9186} (\bibinfo {year} {2020})}\BibitemShut {NoStop}%
\bibitem [{\citenamefont {Nehra}\ \emph {et~al.}(2022)\citenamefont {Nehra}, \citenamefont {Sekine}, \citenamefont {Ledezma}, \citenamefont {Guo}, \citenamefont {Gray}, \citenamefont {Roy},\ and\ \citenamefont {Marandi}}]{Nehra2022FewCycleSqueezing}%
  \BibitemOpen
  \bibfield  {author} {\bibinfo {author} {\bibfnamefont {R.}~\bibnamefont {Nehra}}, \bibinfo {author} {\bibfnamefont {R.}~\bibnamefont {Sekine}}, \bibinfo {author} {\bibfnamefont {L.}~\bibnamefont {Ledezma}}, \bibinfo {author} {\bibfnamefont {Q.}~\bibnamefont {Guo}}, \bibinfo {author} {\bibfnamefont {R.~M.}\ \bibnamefont {Gray}}, \bibinfo {author} {\bibfnamefont {A.}~\bibnamefont {Roy}},\ and\ \bibinfo {author} {\bibfnamefont {A.}~\bibnamefont {Marandi}},\ }\href {https://doi.org/10.1126/science.abo6213} {\bibfield  {journal} {\bibinfo  {journal} {Science}\ }\textbf {\bibinfo {volume} {377}},\ \bibinfo {pages} {1333} (\bibinfo {year} {2022})}\BibitemShut {NoStop}%
\bibitem [{\citenamefont {Tasker}\ \emph {et~al.}(2021)\citenamefont {Tasker}, \citenamefont {Frazer}, \citenamefont {Ferranti}, \citenamefont {Allen}, \citenamefont {Brunel}, \citenamefont {Tanzilli}, \citenamefont {D'Auria},\ and\ \citenamefont {Matthews}}]{Tasker2021GHzHomodyne}%
  \BibitemOpen
  \bibfield  {author} {\bibinfo {author} {\bibfnamefont {J.~F.}\ \bibnamefont {Tasker}}, \bibinfo {author} {\bibfnamefont {J.}~\bibnamefont {Frazer}}, \bibinfo {author} {\bibfnamefont {G.}~\bibnamefont {Ferranti}}, \bibinfo {author} {\bibfnamefont {E.~J.}\ \bibnamefont {Allen}}, \bibinfo {author} {\bibfnamefont {L.~F.}\ \bibnamefont {Brunel}}, \bibinfo {author} {\bibfnamefont {S.}~\bibnamefont {Tanzilli}}, \bibinfo {author} {\bibfnamefont {V.}~\bibnamefont {D'Auria}},\ and\ \bibinfo {author} {\bibfnamefont {J.~C.~F.}\ \bibnamefont {Matthews}},\ }\href {https://doi.org/10.1038/s41566-020-00715-5} {\bibfield  {journal} {\bibinfo  {journal} {Nature Photonics}\ }\textbf {\bibinfo {volume} {15}},\ \bibinfo {pages} {11} (\bibinfo {year} {2021})}\BibitemShut {NoStop}%
\bibitem [{\citenamefont {Bourassa}\ \emph {et~al.}(2021)\citenamefont {Bourassa}, \citenamefont {Quesada}, \citenamefont {Tzitrin}, \citenamefont {Sz{\'a}va}, \citenamefont {Isacsson}, \citenamefont {Izaac}, \citenamefont {Sabapathy}, \citenamefont {Dauphinais},\ and\ \citenamefont {Dhand}}]{Bourassa2021}%
  \BibitemOpen
  \bibfield  {author} {\bibinfo {author} {\bibfnamefont {J.~E.}\ \bibnamefont {Bourassa}}, \bibinfo {author} {\bibfnamefont {N.}~\bibnamefont {Quesada}}, \bibinfo {author} {\bibfnamefont {I.}~\bibnamefont {Tzitrin}}, \bibinfo {author} {\bibfnamefont {A.}~\bibnamefont {Sz{\'a}va}}, \bibinfo {author} {\bibfnamefont {T.}~\bibnamefont {Isacsson}}, \bibinfo {author} {\bibfnamefont {J.}~\bibnamefont {Izaac}}, \bibinfo {author} {\bibfnamefont {K.~K.}\ \bibnamefont {Sabapathy}}, \bibinfo {author} {\bibfnamefont {G.}~\bibnamefont {Dauphinais}},\ and\ \bibinfo {author} {\bibfnamefont {I.}~\bibnamefont {Dhand}},\ }\href {https://doi.org/10.1103/PRXQuantum.2.040315} {\bibfield  {journal} {\bibinfo  {journal} {PRX Quantum}\ }\textbf {\bibinfo {volume} {2}},\ \bibinfo {pages} {040315} (\bibinfo {year} {2021})}\BibitemShut {NoStop}%
\bibitem [{\citenamefont {Guseynov}\ \emph {et~al.}(2026)\citenamefont {Guseynov}, \citenamefont {Holmes},\ and\ \citenamefont {Angrisani}}]{Guseynov2026CoherentStatePropagation}%
  \BibitemOpen
  \bibfield  {author} {\bibinfo {author} {\bibfnamefont {N.}~\bibnamefont {Guseynov}}, \bibinfo {author} {\bibfnamefont {Z.}~\bibnamefont {Holmes}},\ and\ \bibinfo {author} {\bibfnamefont {A.}~\bibnamefont {Angrisani}},\ }\href@noop {} {\bibinfo {title} {Coherent-state propagation: A computational framework for simulating bosonic quantum systems}} (\bibinfo {year} {2026}),\ \Eprint {https://arxiv.org/abs/2604.19625} {arXiv:2604.19625 [quant-ph]} \BibitemShut {NoStop}%
\end{thebibliography}%

\end{document}